%% file: bayes.tex
\documentclass[useAMS,usenatbib]{mn2e}

\usepackage{graphicx}
\usepackage{amsmath}
\usepackage{dcolumn}

\title[Stellar magnetic parameters from Bayesian analysis]{Stellar magnetic field parameters from a Bayesian analysis of high-resolution spectropolarimetric observations\thanks{Based on observations obtained at the Canada-France-Hawaii Telescope (CFHT) and at the T\'elescope Bernard Lyot (TBL). CFHT is operated by the National Research Concil of Canada, the Institut National des Sciences de l'Univers of the Centre National de la Recherche Scientifique of France, and the University of Hawaii. TBL is operated by CNRS/INSU.}}

\author[V. Petit \& G.~A. Wade]{V. Petit$^{1}$\thanks{E-mail: VPetit@wcupa.edu}, 
and G.A. Wade$^{2}$
\\
$^{1}$Dept. of Geology \& Astronomy, West Chester University, West Chester, PA 19383\\
$^{2}$Dept. of Physics, Royal Military College of Canada, PO Box 17000, Stn Forces, Kingston, Canada, K7K 4B4}

\begin{document}
\include{aas_macros}

\date{Accepted 2011 October 26.  Received 2011 October 22; in original form 2011 August 17}

\pagerange{\pageref{firstpage}--\pageref{lastpage}} \pubyear{2010}

\maketitle

\label{firstpage}

\begin{abstract}
In this paper we describe a Bayesian statistical method designed to infer the magnetic properties of stars observed using high-resolution circular spectropolarimetry in the context of large surveys. This approach is well suited for analysing stars for which the stellar rotation period is not known, and therefore the rotational phases of the observations are ambiguous. 
The model assumes that the magnetic observations correspond to a dipole oblique rotator, a situation commonly encountered in intermediate and high-mass stars. Using reasonable assumptions regarding the model parameter prior probability density distributions, the Bayesian algorithm determines the posterior probability densities corresponding to the surface magnetic field geometry and strength by performing a comparison between the observed and computed Stokes $V$ profiles. 

Based on the results of numerical simulations, we conclude that this method yields a useful estimate of the surface dipole field strength based on a small number (i.e. 1 or 2) of observations. On the other hand, the method provides only weak constraints on the dipole geometry. The odds ratio, a parameter computed by the algorithm that quantifies the relative appropriateness of the magnetic dipole model versus the non-magnetic model, provides a more sensitive diagnostic of the presence of weak magnetic signals embedded in noise than traditional techniques.

To illustrate the application of the technique to real data, we analyse seven ESPaDOnS and Narval observations of the early B-type magnetic star LP\,Ori. Insufficient information is available to determine the rotational period of the star and therefore the phase of the data; hence traditional modelling techniques fail to infer the dipole strength. In contrast, the Bayesian method allows a robust determination of the dipole polar strength, $B_d=911^{+138}_{-244}$\,G.
\end{abstract}

\begin{keywords}
stars: magnetic fields-- stars: early-type -- techniques: polarimetric -- methods: statistical.
\end{keywords}

\section{Introduction}

The evolution of a massive star is strongly determined by its rotation, as well as the mass lost through its stellar wind, both of which can be strongly influenced by the presence of a magnetic field \citep[e.g.][]{2010ApJ...714L.318T,2011arXiv1106.3008W}. A magnetic field can furthermore couple different layers of a star's interior, thereby modifying internal differential rotation and circulation currents \citep{2005A&A...440.1041M}. If a field has a large-scale component that extends outside the stellar surface, it can also channel the outflowing stellar wind, creating a structured wind - a magnetosphere - which will modify the rate and geometry of mass loss along with its observational properties \citep{1978ApJ...224L...5L,1990ApJ...365..665S,1997ApJ...485L..29B,1997A&A...323..121B}. Furthermore, if the field couples the rotating surface of the star with the outflowing stellar wind, both effects will result in angular momentum loss (via the outflowing stellar wind), which differs strongly from that of a non-magnetic star \citep{2009MNRAS.392.1022U}. As angular momentum and mass loss are cornerstone inputs in stellar evolution calculations, it is crucial that the effect of magnetic fields in massive stars be understood properly. For example,  evolutionary tracks and isochrones can be used to interpret large datasets associated with OB associations like the \textit{VLT-FLAMES} surveys of massive stars \citep[][]{2008A&A...479..541H}.
 
Over the last decade, our knowledge of the properties of massive star magnetic fields has significantly improved, in large part due to a new generation of powerful high-resolution spectropolarimetric instrumentation. Traditionally, stellar magnetic fields were modelled using measurements of the longitudinal magnetic field of the star, yielding a single quantity: the strength of the disc-integrated line-of-sight component of the magnetic field \citep[e.g.][]{2006A&A...450..777B}. In the absence of a field detection, the interpretation of such data was entirely dependent on the (unknown) stellar and magnetic geometry, and therefore highly ambiguous. However, modern high-resolution spectropolarimetry yields Doppler-broadened, velocity-resolved Stokes profiles measured across spectral lines, which encode additional information about the field geometry. These data therefore allow a more robust inference of the field characteristics.

Remarkably detailed information about the strength and local/global structure of stellar magnetic fields can be extracted from high-resolution, phase-resolved spectropolarimetric datasets acquired in two or four Stokes parameters (i.e. Stokes $IV$ or Stokes $IVQU$) using techniques such as Magnetic Doppler Imaging \citep[e.g.][]{2010A&A...513A..13K}. However, detailed modelling of this sort relies on extensive high-quality datasets that demand significant telescope time, and which can only be obtained for a small number of stars. The lower-quality data that can be obtained for stars with less suitable observational properties can still be approximately investigated using parametric models; nevertheless, even such a simple approach often requires approximately a dozen observations per star.

Large observing programs, such as the Magnetism in Massive Stars (MiMeS) project, have dedicated significant resources to survey large samples of massive stars in search of magnetic fields \citep{2010arXiv1009.3563W}. Such surveys typically acquire a small number of Stokes $V$ observations (typically 1-3) per star, for a large number of stars. An outstanding problem concerns how to extract useful information about the magnetic field of a star from such a limited data set. 

In this paper we describe an approach to constrain the magnetic field strength of a star from small numbers of high-resolution Stokes $V$ observations, assuming the dipole oblique rotator paradigm, and expressed using the formalism of Bayesian statistics. 
In Sect. \ref{sec|line_synth} we describe briefly the model used to synthesise the emergent local circular polarisation produced by the Zeeman effect under the weak-field approximation. We present the disk-integrated emergent Stokes $V$ profiles obtained for a dipolar magnetic topology of a rotating star. In Sect. \ref{sec|bayes} we introduce the Bayesian algorithm used to compare a set of high-resolution Stokes $V$ observations with a grid of synthetic line profiles. 
Sect. \ref{sec|test} presents the application of the Bayesian algorithm to simulated data. 
We demonstrate that the Bayesian odds ratio can help detect a weak magnetic signal, by quantifying which target should be re-observed. It can also discriminate, with a few additional observations, whether an observation has a noise pattern that by chance looks like a magnetic signal versus a real magnetic signal. We also show that the Bayesian algorithm provides a meaningful upper limit on the magnetic strength  in the case of a non-detection, and is able to reliably estimate the dipole field strength in the case of a detection. Some limited constraints can also be obtained for the geometry of the dipole.
We compare the Bayesian diagnostics with the traditional global longitudinal field and signal detection probability diagnostics. 
Finally, in Sect. \ref{sec|lpori}, we present the Bayesian results for the magnetic B-type star LP\,Ori \citep{2008MNRAS.387L..23P}, obtained with observations of various signal-to-noise ratios.

\section{Line synthesis}
\label{sec|line_synth}

\subsection{Local profiles}

\label{sec|model}

Splitting of spectral lines due to the Zeeman effect corresponds to about 1\,km\,s$^{-1}$ per kG of surface field modulus. This implies that even relatively strong fields are difficult to detect reliably from Zeeman splitting in the spectrum of most stars. The situation is particularly challenging in the case of hot stars, where thermal broadening (of order a few km\,s$^{-1}$), turbulent broadening (of order a few tens of km\,s$^{-1}$) and rotational broadening (potentially a few hundreds of km\,s$^{-1}$) combine to fully obscure any modification of the line profile due to the Zeeman effect.

The Zeeman effect provides us with a second useful tool, in the form of the polarisation of the Zeeman components. Zeeman components exist in two different types: $\pi$ components, spread symmetrically about the zero-field wavelength of the spectral line and $\sigma$ components, with symmetric wavelength offsets to the red and blue of the zero-field wavelength. If the external magnetic field is oriented parallel to the observer's line of sight (a longitudinal field), the $\pi$ components vanish, while the two groups of $\sigma$ components have opposite circular polarisations. In a field aligned perpendicular to the line of sight (a transverse field), the $\pi$ components are linearly polarised perpendicular to the field direction, while the $\sigma$ components are linearly polarised parallel to the field direction. Therefore, spectropolarimetric observations are an extremely useful tool for detection and characterisation of both the strength and orientation of stellar magnetic fields. 

In order to correctly predict the polarised spectrum that will emerge from a stellar atmosphere, one must perform radiative transfer in an anisotropic medium, where the propagation properties depend on the propagation directions. 
In the weak-field regime, we can treat the anisotropic absorption and dispersion profiles as perturbations of the isotropic case \citep{2004ASSL..307.....L}. In that case, at first order, only the longitudinal component of the magnetic field contributes to the polarisation, and therefore only circular polarisation is predicted. The contribution of the transverse field, and the occurrence of linear polarisation, is a second-order effect.
Therefore, circular polarisation (Stokes $V$ spectra) signatures are generally stronger, and used for large surveys as such observations provide the best detection threshold.

Although there are some elaborate codes that are able to accurately synthesise the detailed circular polarisation profile of an arbitrary spectral transition \citep[for a review see][]{2001A&A...374..265W}, we will use the weak-field approximation, in order to draw a simple picture of the Stokes $V$ spectrum emerging from a star. 
The 1st order solution for the circular polarisation to the polarised radiative transfer equations is given by:
	\begin{equation}
		V(v) = -\frac{e}{4\pi mc^2}cg_\mathrm{eff}\lambda_0B_{||}\frac{dI(v)}{dv}.\label{eq|stokesV}
	\end{equation}
The Stokes $V$ profile $V(v)$ emerging from a point at the surface of a star (referred to as a \textit{local} profile) has the same shape as the derivative of the local intensity profile $I(v)$, scaled by the longitudinal component of the magnetic field $B_{||}$ at that point, by the wavelength $\lambda_0$ of the transition, and by the effective Land\'e factor $g_\mathrm{eff}$ of the transition. The effective Land\'e factor represents the separation of a triplet Zeeman pattern that approximates a more complex Zeeman pattern, and can be calculated under LS coupling, or taken from atomic experiments. Although we will be using the weak field approximation throughout this paper, the Bayesian algorithm can use synthetic profiles computed with any spectrum synthesis code.

A multi-line approach will usually be applied -- in order to increase the signal to noise ratio (s/n) -- to the data for which a Bayesian approach will be useful. Although these multi-line techniques have been developed and refined for more than two decades now \citep[e.g.][]{1981ApJ...247..569B,1996SolarPhys...164...417S}, the most widely-used approach is the Least Squares Deconvolution (LSD) method introduced by \citet{1997MNRAS.291..658D}. Under the weak field approximation, the LSD method assumes that all the spectral lines have the same shape, and that this shape is scaled by the line depth for the intensity line profiles (Stokes $I$) and by the product of the line depth, the wavelength and the effective Land\'e factor for the circularly polarised line profile (Stokes $V$). The LSD method therefore already approximates the Zeeman pattern of a line by a triplet whose separation is given by an effective Land\'e factor. For a complete discussion on the circumstances under which a LSD profile can be treated as a single spectral line, see \citet{2010A&A...524A...5K}.

\subsection{Disk integration}
 \label{sec|disk}
 
In order to predict the Stokes $V$ spectrum of a given stellar spectral line, we must take into account the contribution from every point on the visible hemisphere. In this paper, we consider a single spectral line of arbitrary depth and a width corresponding to the sum of the spectral resolution and an ad hoc local broadening width. The contribution of each point on the visible hemisphere is dictated by the effective surface area and a wavelength-independent limb-darkening of the form:
\begin{equation}
	\frac{I_c}{I_{c,0}} = 1-\epsilon + \epsilon\cos\theta,	
\end{equation}
where $I_{c,0}$ is the intensity at the stellar disk centre, $\theta$ is the angle between the normal to the surface and the line of sight, and $\epsilon$ is the limb-darkening factor \citep{1992oasp.book.....G}. We adopt $\epsilon=0.6$ for the rest of this paper. 
The surface velocity field of the stellar disk is defined by the projected equatorial velocity $v\sin i$ assuming rigid rotation. 

We consider a simple magnetic topology with a dipolar field of strength $B_p$, as the vast majority of intermediate mass and high mass stars' magnetic fields appear to be predominantly dipolar \citep[e.g.][]{1980ApJS...42..421B,2007A&A...475.1053A}. The orientation of the dipole with respect to the observer's reference frame is described by the inclination $i$ of the rotational axis to the line of sight, the rotational phase $\varphi$ and the obliquity $\beta$ of the magnetic axis with respect to the rotational axis.

\begin{figure}
\includegraphics[width=75mm]{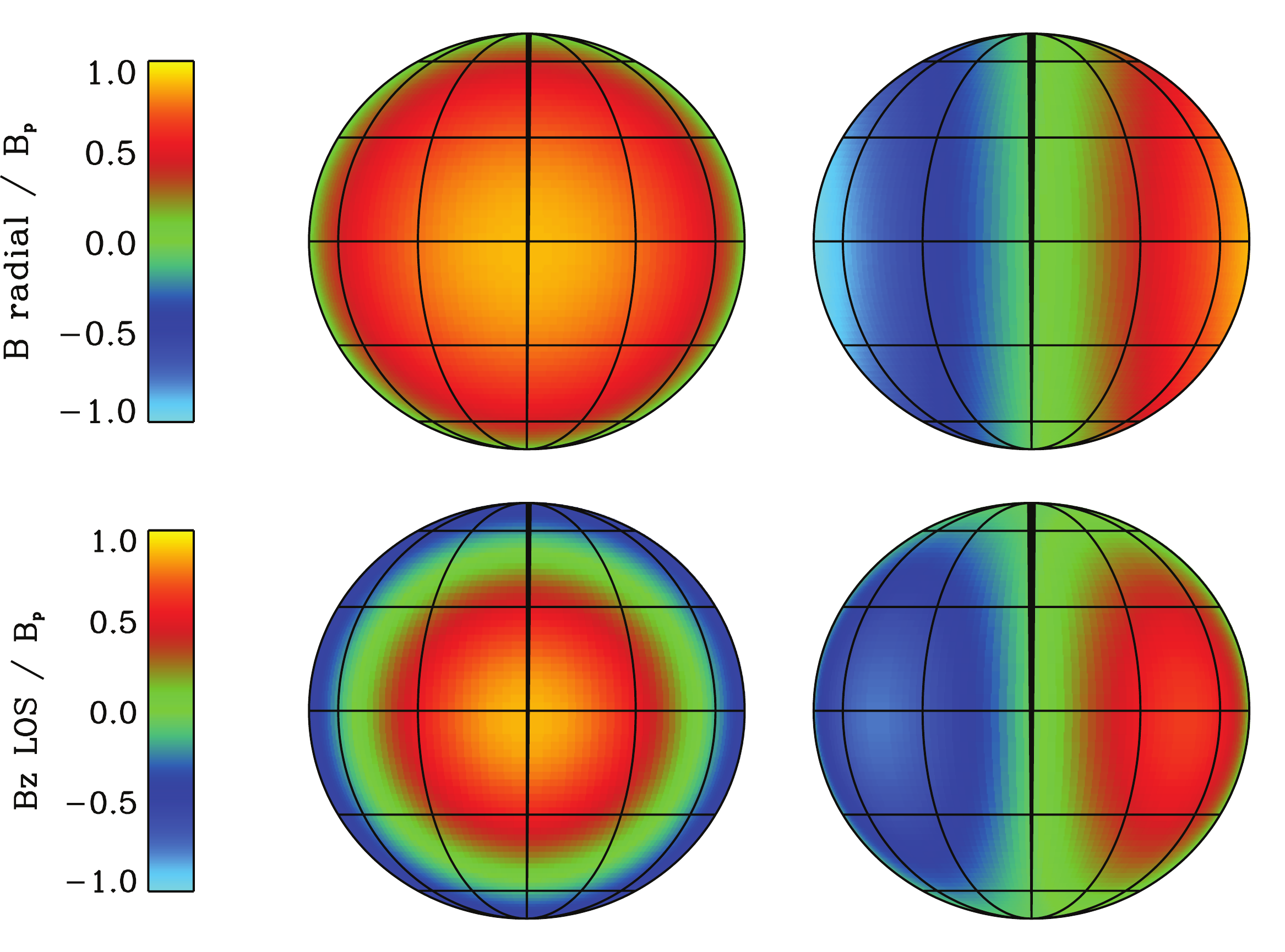}
\caption{\label{fig|model_bz} Radial (top) and longitudinal (bottom) surface magnetic field for a dipolar field seen pole-on (left) and equator-on (right). For the radial field, red, blue and green colours represent field oriented away from, into, and parallel to the star surface, respectively. For the longitudinal field, red, blue and green colours represent fields oriented toward, away from and perpendicular to the observer's line of sight, respectively. The black grid represent the rotation reference frame, with an inclination $i=90^\circ$. The obliquity of the field is $\beta=90^\circ$ and the pole-on and equator-on configurations would correspond to rotational phases $\varphi=0^\circ$ and $\varphi=90^\circ$, respectively.}
\end{figure}

Figure \ref{fig|model_bz} shows the radial field (top) and the longitudinal field (bottom) for a dipole field seen positive pole-on (left) and magnetic equator-on (right). 
The radial field is defined by its orientation with respect to the stellar surface normal. For example, when we are looking at the positive pole (left), the radial field is at its maximum (red) in the middle of the stellar disk. On the edge of the disk, the field is parallel to the surface, and the radial field is null (green). 

The longitudinal field is the important quantity for the polarised radiative transfer. The longitudinal field represents the projection of the magnetic field vectors with respect to the observer's line of sight (perpendicular to the paper plane). If we look at the magnetic field seen pole-on (bottom left), all of the magnetic vectors that are oriented toward the observer are colour-coded in red shades, green corresponds to a null longitudinal field component, and magnetic vectors oriented away from the observer are colour-coded in blue shades. As seen in Eq. (\ref{eq|stokesV}), two longitudinal magnetic fields of same magnitude but opposite signs will produce inverse local Stokes $V$ profiles. However, a good fraction of the field components oriented away from the observer are located on the hidden hemisphere of the star. There is therefore a net positive longitudinal component on the visible hemisphere (i.e. the visible hemisphere is more red than blue). In terms of the total emergent Stokes $V$, positive local profiles (local profiles for which the longitudinal field is positive) will dominate the total line profile. This effect is even more pronounced when limb darkening is taken into account, as the edges of the disk contribute less to the total brightness.

We now turn our attention to the case where we are looking at the magnetic equator (bottom right). In principle, there are as many magnetic vectors pointing toward the observer as away. The global longitudinal component is therefore null, as every magnetic vector on the left side of the stellar disk has its opposite on the right side of the disk. In terms of the emerging Stokes $V$ profile, each positive local Stokes $V$ profile will be cancelled out by the negative local Stokes $V$ profile of same amplitude coming from the opposite side of the visible stellar disk, even when considering limb-darkening. The resulting Stokes $V$ profile should therefore be null when the global longitudinal component is null. This is effectively the case, if the star is not rotating. 

If the star is rotating around a rotation axis oriented toward the top of the paper (as represented by the black grid), the line of sight component of that rotation will introduce some Doppler shifts to different points on the stellar surface. For example, if the star rotates from left to right, the point situated to the extreme left of the stellar disk will be shifted by $-v\sin i$. The point situated at the extreme right of the stellar disk, whose local Stokes $V$ profile would in principle cancel out the one produced by the leftmost point, will now be shifted by $+v\sin i$. Therefore, the rotation is able to separate in the velocity space the local Stokes $V$ line profiles that would otherwise cancel out, and a net circular polarisation profile can be seen even if the global longitudinal field is null. It has been shown that a $v\sin i$ as low as a few km\,s$^{-1}$ is enough for instruments able to resolve this effect in the line profile \citep{2000MNRAS.313..823W}. 

\begin{figure}
\includegraphics[width=84mm]{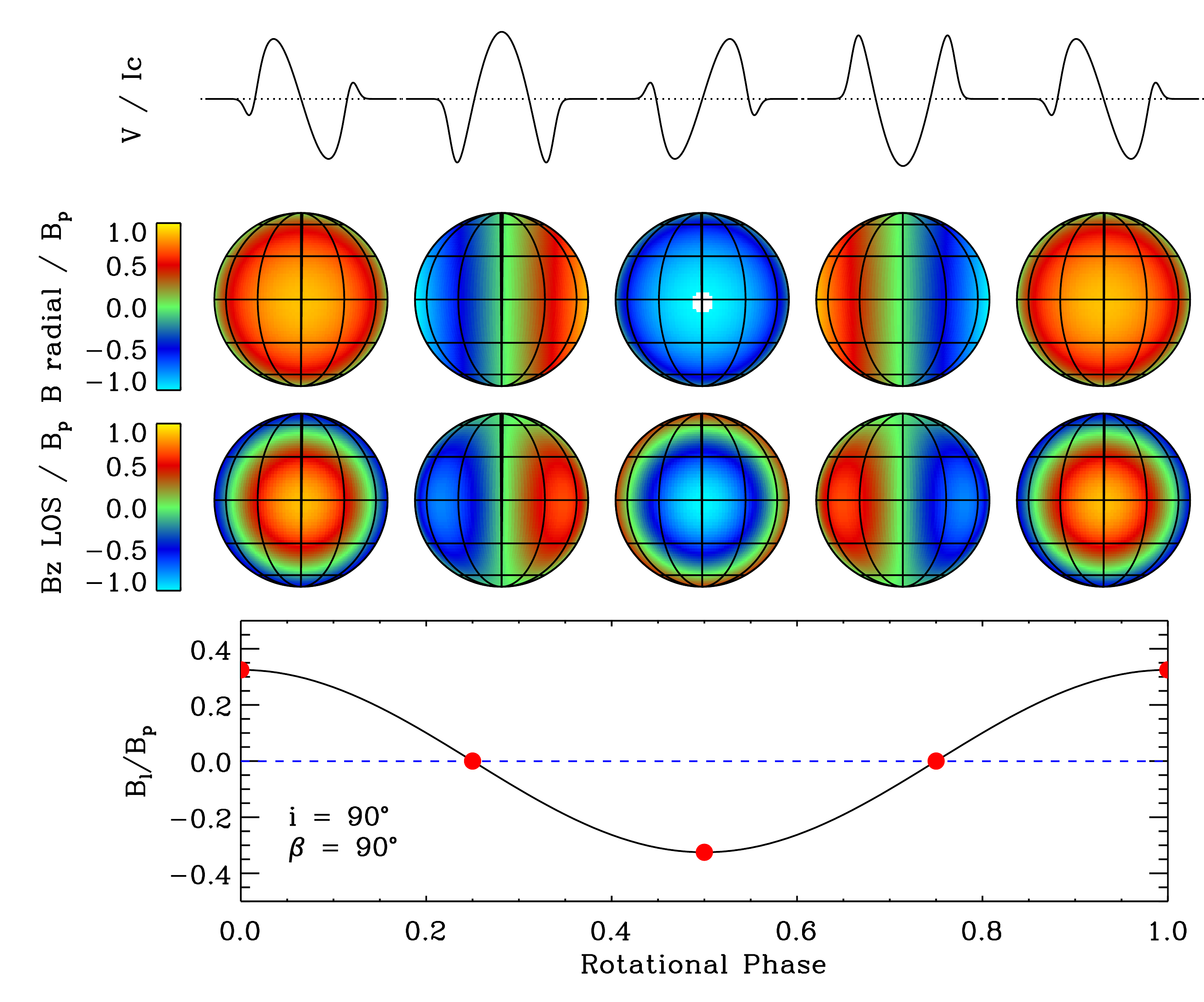}\\
\caption{\label{fig|model_nodetect1_cont0} \textit{Top:} Shape of the Stokes $V$ line profiles for five rotational phases of a star with a dipolar field with $i=\beta=90^\circ$. \textit{Middle:} Corresponding radial field and longitudinal field components (colour schemes are as indicated in Figure \ref{fig|model_bz}). The rotation axis inclination $i=90^\circ$ is represented by the black grid and the magnetic field pole is perpendicular to the rotation axis ($\beta=90^\circ$). \textit{Bottom:} Global longitudinal field curve, normalised to the dipole field strength. The dots show the phase of the profiles. Note how the Stokes $V$ profile do not disappear when the global longitudinal field goes to zero.}
\end{figure}

Figure \ref{fig|model_nodetect1_cont0} shows an example of the shapes and relative amplitudes of Stokes $V$ profiles emerging from a star rotating at 50\,km\,s$^{-1}$ (top). The radial and longitudinal fields are shown (middle), as well as the global longitudinal field curve (bottom). The phases indicated by red dots on the curve indicate the rotational phases corresponding to each shown profile. We can see that for this dipole configuration ($i=90^\circ$, $\beta=90^\circ$), the emerging Stokes $V$ profile has an amplitude as strong when the global longitudinal field is null (2nd and 4th phases) than when it is at its maximum (1st, 3rd and last phases). We note the Stokes $V$ profiles at $B_l=0$\,G are symmetric around the centre of the line. 
If we were to observe such profiles with an instrument that does not resolve the line profiles, we would indeed obtain a null global longitudinal field, even if the profiles are as clearly detectable with high-resolution observations than when the longitudinal field is at its maximum. 

\begin{figure}
\includegraphics[width=84mm]{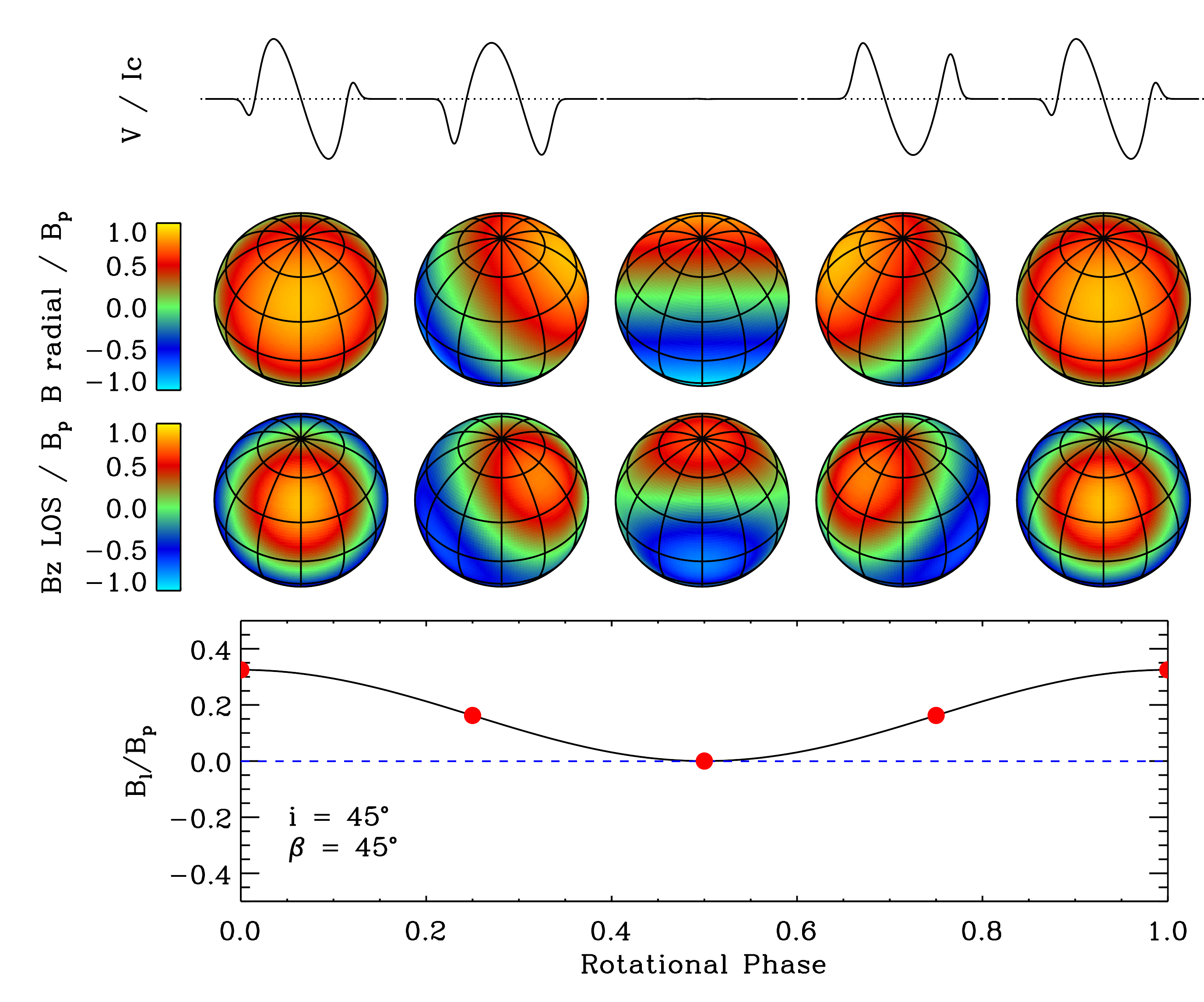}
\caption{\label{fig|model_nodetect1_cont2} Same as Figure \ref{fig|model_nodetect1_cont0} for a rotation axis inclined by $i=45^\circ$ and a magnetic pole of obliquity $\beta=45^\circ$. Note how the Stokes $V$ signal disappears when the global longitudinal field goes to zero because $i+\beta\sim90^\circ$.  }
\end{figure}

Figure \ref{fig|model_nodetect1_cont2} shows a second dipole configuration ($i=45^\circ$, $\beta=45^\circ$), where the rotation is not able to separate components that will cancel out. When the longitudinal field is null (middle profile, we are looking at the magnetic equator), we can see that as the star rotates from left to right, all the points situated on the left side of the stellar disk will be shifted to the blue. However, each half of the visible hemisphere contains as many field vectors oriented toward the observer as oriented away from the observer. Therefore, all the red-shaded points situated on the left side of the stellar disk will have a corresponding inverse profile that will be Doppler shifted by the same amount, and the cancellation will occur. This particular symmetry occurs when $\beta+i\sim90^\circ$, at a phase where we are looking at the magnetic equator. 

Due to the intrinsic symmetry of a dipole field, as well as the fact that circular polarisation is only sensitive to the longitudinal field components, some parameter degeneracy is encountered for the resulting Stokes $V$ profiles. For example, if a dipolar field of parameter $i$, $\beta$ and $\varphi$ produces a profile $f$, the following symmetries (or anti-symmetries) occur:
\begin{equation}
	\label{eq|model_sym}
	\begin{pmatrix} 90^\circ-i \\ 90^\circ-\beta \\ \varphi \\ f \end{pmatrix}, 
	\begin{pmatrix} 90^\circ-i \\ \beta \\ 180^\circ+\varphi \\ -f \end{pmatrix}, 
	\begin{pmatrix} i \\ 90^\circ-\beta \\ 180^\circ+\varphi \\ -f \end{pmatrix}.
\end{equation}

\section{The bayesian algorithm}
\label{sec|bayes}

Bayesian statistics have proven to be an useful tool to find plausible models to explain astronomical data \citep[e.g.][]{2011A&A...532A.116T,2011ApJ...740...44A}. 
A probability density distribution $p(H_i|I)$ provides a quantitative encoding of the plausibility of a certain hypothesis ($H_i$), given our current state of knowledge ($I$). The main pathway to Bayesian statistics \citep{Jaynes2003} is the Bayes theorem:
\begin{equation}
	p(H_i|D,I) = \frac{p(H_i|I)p(D|H_i,I)}{p(D|I)}.
\end{equation}
This theorem states that the posterior probability density $p(H_i|D,I)$ of an hypothesis, given a new dataset $D$ and the current state of knowledge, is equal to the product of prior knowledge of the plausibility of the hypothesis $p(H_i|I)$ and the likelihood $p(D|H_i,I)$ of obtaining the new dataset if the hypothesis is true. The global likelihood, $p(D|I)=\sum_i p(H_i|I)p(D|H_i,I)$, acts as a normalisation constant. The act of summing (or integrating in the case of a continuous hypothesis set) is called marginalisation.

One difficulty of modelling the Stokes $V$ spectra of stars observed as part of a survey like MiMeS is that most of the time the rotational phase of the star at the moment of the observation is not known, because the rotational period is generally unknown. Our approach is therefore to use Bayesian probability nomenclature to find which set of phase-independent configurations $\mathcal{B} = (B_p, i, \beta)$ can reproduce the observations. In other words, while the  geometry $\mathcal{B}$ of the field must stay the same for each observation $n$ of a given star, the phase $\varphi_n$ may in principle have any value. 

The hypothesis to be tested is the presence of an oblique dipolar magnetic field in a particular star. The predictions of this hypothesis are represented by the model $M_1$, parametrized with the parameters $\mathcal{B}$ (=[$B_p$, $i$, $\beta$]) and $\Phi$ (=[$\varphi_1$..$\varphi_N$]), the latter representing a set of phases associated with a set of Stokes $V$ observations $\mathcal{D}$ (=[$D_1$..$D_N$]).
The Bayes theorem tells us that the joint posterior probability density for the parameters, assuming the veracity of the model $M_1$, is
\begin{equation} \label{eq|post_conf}
	p(\mathcal{B}, \Phi | \mathcal{D}, M_1) = \frac{p(\mathcal{B}, \Phi | M_1)p(\mathcal{D} | \mathcal{B}, \Phi, M_1)}{p(\mathcal{D} | M_1)}. 
\end{equation}
Our prior knowledge of the probability density for each model parameter, which can be as simple as its expected range, can be encoded in the prior term $p(\mathcal{B}, \Phi | M_1)$, whereas the information brought forward by the new observations is reflected in the likelihood term $p(\mathcal{D} | \mathcal{B}, \Phi, M_1)$. The global likelihood is the normalisation term $p(\mathcal{D} | M_1)$, which is equal to the total probability:
\begin{equation} \label{eq|globalLH}
	p(\mathcal{D} | M_1)=\int\int p(\mathcal{B}, \Phi | M_1)p(\mathcal{D} | \mathcal{B}, \Phi, M_1) \mathrm{d}\Phi \mathrm{d}\mathcal{B}.
\end{equation}
We then treat the set of phases $\Phi$ as nuisance parameters by marginalising the joint posterior probability:
\begin{equation}  \label{eq|post_conf_b}
	p(\mathcal{B} | \mathcal{D}, M_1) = \int p(\mathcal{B}, \Phi | \mathcal{D}, M_1)  \mathrm{d}\Phi.
\end{equation}
We therefore ensure that an adequate $\mathcal{B}$ possesses Stokes $V$ profiles at some phases that match all the observations. From this $\mathcal{B}$ posterior probability density, we can then explore the plausible region of the parameter space of an oblique dipole field given the data, assuming that the model is true.

In the case of a limited number of spectropolarimetric observations, we will be generally interested in the field strength value, as only weak constraints can be placed on the geometrical parameters. This is particularly relevant in the case of a non-detection, where we wish to put an upper limit on the strength of an undetected dipole field. To obtain the posterior probability density for a given parameter, we need to marginalise the joint probability $p(\mathcal{B}|\mathcal{D}, M_1)$ over the other parameters. 
For example, the posterior probability density marginalised for the field strength is:
\begin{equation}
	p(B_p|\mathcal{D}, M_1) = \int p(\mathcal{B}|\mathcal{D}, M_1)\,\mathrm{d}i\,\mathrm{d}\beta .
\end{equation}

Another powerful application of Bayesian statistics is the ability to quantitatively test the plausibility of one hypothesis versus another. We can therefore compare the plausibility of the oblique dipole model $M_1$, by computing the so-called odds ratio of this model compared to the model $M_0$ representing the absence of a magnetic field. To do so, we need to compute the posterior probability $p(M_i|D,I)$ of a given model which is, according to the Bayes theorem:
\begin{equation}
	p(M_i|D,I) = \frac{ p(M_i|I)p(D|M_i, I) }{ p(D|I) }.
\end{equation}
The prior term $p(M_i|I)$ encodes the plausibility of the model, given our current knowledge, and the global likelihood $p(D|M_i, I)$ encodes the plausibility of the model, given the new observations. 

Therefore, the odds ratio of our two competing models, $M_0$ and $M_1$, can be written as:
\begin{equation}
	\mbox{odds ($M_0/M_1$)} = \frac{p(M_0|\mathcal{D},I)}{p(M_1|\mathcal{D},I)} = \frac{p(M_0|I)}{p(M_1|I)} \frac{p(\mathcal{D}|M_0, I)}{p(\mathcal{D}|M_1, I)}.
\end{equation}
Typically, as no model is preferred prior to the acquisition of Stokes $V$ observations, the ratio of priors $p(M_0|I)/p(M_1|I)$ will be set to unity. The global likelihood of the dipole model is given by Eq. (\ref{eq|globalLH}). As the model representing the absence of magnetic field has no parameters, its global likelihood is simply the product of the likelihoods that $V=0$ for each observation. 
According to \citet{Jeffreys1998}, the evidence in favour of a model is considered weak when the odds are $>10^{0.5}$ ($\sim$3:1), moderate when $>10^{1}$ ($\sim$10:1), strong when $>10^{1.5}$ ($\sim$30:1) and very strong when $>10^2$ ($\sim$100:1).

\subsection{Practical implementation}

Given the complexity of disk integration for a rotating magnetic star, it is not practical to solve the Bayesian problem analytically \citep[for an example applied to the local Stokes profiles of the Sun, see][]{2011ApJ...731...27A}. 
A Markov-chain Monte-Carlo method works by choosing series of parameter sets, the parameter regions with a high likelihood being more likely to be picked. The sampling itself therefore reflects the posterior probability density. This method is not well suited to this problem because a large number of calculations are required to assess the probability of each $\mathcal{B}$ configuration, due to the marginalisation over all possible phases. 
We therefore chose a numerical, brute-force approach in order to explore the parameter space by sampling it with a regular grid.

When evaluating Eqs. (\ref{eq|post_conf}) and (\ref{eq|post_conf_b}), we can make use of the fact that the prior probability densities for each parameter are independent such that $p(\mathcal{B}, \Phi | M_1) = p(\mathcal{B}|M_1)p(\Phi|M_1)$. 
Furthermore, the prior probability for the phases are the same such that $p(\varphi_n)=p(\varphi)$. Finally, the likelihood $p(\mathcal{D}|\mathcal{B}, \Phi, M_1)$ can be factored into $\Pi_{n=1}^N p(D_n|\mathcal{B}, \varphi_n, M_1)$ given that the likelihood of one observation $D_n$ is only dependent on the phase $\varphi_n$. Therefore, Eq. \ref{eq|post_conf_b} can be written as:

\begin{equation}
	\label{eq|post_conf_new}
	p(\mathcal{B}|D,M_1)=p(\mathcal{B}|M_1)\frac{ \prod_{n=1}^N \int p(\varphi|M_1)p(D_n|\mathcal{B},\varphi,M_1){\rm d}\varphi}{p(\mathcal{D}|M_1)}.
\end{equation}

For each point in a [$B_p$, $i$, $\beta$, $\varphi$] parameter space, we can compute the likelihood between the synthetic profile produced by these parameters and each observation. In fact, given the Stokes $V$ profile symmetry properties described in Eq. (\ref{eq|model_sym}), only the synthetic profiles for a quarter of the parameter space need to be computed. Furthermore, synthetic profile interpolation is possible in the $B_p$ direction, as the local Stokes $V$ profile amplitude increases linearly although the shape remains the same.
The maginalisation integrals are performed numerically using a five point Newton-Cotes algorithm. An adequate sampling of the $M_1$ parameter space was chosen to obtain accurate values of $p(M_1|\mathcal{D},I)$.

\subsection{Probability densities}

For each parameter of the oblique dipole model, a prior probability density is needed to evaluate Eq. (\ref{eq|post_conf}). The dipole field strength $B_p$ is a parameter that can vary over several decades (from a few dozen G to a few kG). We therefore used a Jeffreys prior, which sets an equal probability per decade and is therefore scale invariant. We used a modified form \citep{2005ApJ...631.1198G} in order to eliminate the singularity at $B_p=0$:
\begin{equation}
	p(B_p|M_1) = \frac{1}{ (B_p+a) \ln\left(\frac{a+B_{p,\mathrm{max}}}{a}\right)}.
\end{equation}
This prior function behaves as a flat prior when $B_p<a$, and as a Jeffreys prior when $B_p>a$. 
The maximum dipolar field strength $B_{p,\mathrm{max}}$ is adjusted according to the strength of the Stokes $V$ signal and by the quality of the data (for example, if the s/n is lower, a field of a higher strength can hide in the noise in the case of a non-detection).  For this paper, the grid has 250 $B_p$ values. A finer sampling might be required when using a large number of observations, in order to avoid under sampling the joint posterior probability density. We tested the effect of the choice of the $a$ parameter, and we settled using a value about twice the $B_p$ grid step.

The inclination of the rotation axis to the line of sight is a ``position'' parameter (invariant under a shift of zero position), for which a flat prior would be well suited. However, we know that if the orientations of the rotation axes of stars are generally randomly distributed, the probability that the inclination angle is  between $i$ and $i+\mathrm{d}i$ is:
\begin{equation}
	\label{eq|prior_i}
	p(i|M_1) = \frac{1}{2}\sin i.
\end{equation}
The inclination could in principle vary from $0^\circ$ to $180^\circ$. However, if a non-null $v\sin i$ is measured, it is possible to put a lower limit on the inclination by estimating the break-up velocity -- typically around 700\,km\,s$^{-1}$ for a massive OB star. We used a grid of 37 $i$ values, giving a sampling at least every $\sim5^\circ$. 

The same reasoning could apply to the obliquity angle $\beta$ of the magnetic axis to the rotation axis. 
However, we ignore at this point if the magnetic axes are generally randomly distributed, or if a general relation exists for the position of the magnetic axis relative to the rotation axis. 
For example, \citet{2000A&A...359..213L} have shown that the magnetic axis of long-period ApBp stars is generally aligned with their rotation axis. Therefore, instead of using a prior that assumes a randomness of the magnetic axis position, we kept a flat prior for the obliquity angle:
\begin{equation}
	p(\beta|M_1) = \frac{1}{\beta_\mathrm{max}-\beta_\mathrm{min}}.
\end{equation} 
The obliquity varies from $\beta_\mathrm{min}=0^\circ$ to $\beta_\mathrm{max}=180^\circ$, sampled every $5^\circ$.

A flat prior probability is also used for the rotational phase: 
\begin{equation}
	p(\varphi|M_1) = \frac{1}{\varphi_\mathrm{max}-\varphi_\mathrm{min}},
\end{equation} 
with $\varphi_\mathrm{min}=0^\circ$ and $\varphi_\mathrm{max}=360^\circ$, sampled every $5^\circ$.

The likelihood function of a given data set $D_n$ corrupted with Gaussian noise is given by:
\begin{eqnarray}
	p(D_n|\mathcal{B}, \varphi, M_1) = (2\pi)^{-\frac{M}{2}}\left[\prod_{k=1}^M\sigma_k^{-1}\right] \nonumber \\ \exp\left\{-\frac{1}{2}\sum_{k=1}^M\frac{(d_k-f_k)^2}{\sigma_k^2}\right\},\label{eq|LHodds}
\end{eqnarray}
where $d_k$ represents one of the $M$ datapoints with its variance $\sigma_k$, and $f_k$ represents the model prediction.

When performing a parameter estimation, assuming the veracity of a model, it is a good practice to treat the variance of our data as a model parameter itself, using our estimate of the variance (i.e. the error bars) as a starting point. Proceeding in this way, the final results will be less sensitive to our estimation of the variance, as noise propagation can be problematic in heavily processed data, as can be the case for LSD profiles. 
Furthermore, this approach treats any features that cannot be explained by this particular model (for example higher order polar field components or distortions of the profile due to abundance spots) as additional noise. 
According to the maximum entropy principle, a Gaussian distribution will be the most noncommittal about information we do not have, leading to the most conservative estimates of the parameters. If our estimate of the variance of each point is denoted by $s_k$, we introduce a ``noise scaling'' parameter $b$ \citep{2005blda.book.....G}, in order to estimate the total noise variance $\sigma_k$:
\begin{equation}
	\label{eq|prior_b}
	\sigma^2_k=\frac{s^2_k}{b},
\end{equation}
where $b$ varies around unity. 
The resulting expression for the likelihood is:
\begin{eqnarray}
	p(D_n|\mathcal{B}, \varphi, b, M_1)=(2\pi)^{-M/2}(b)^{M/2}\left[\prod_{k=1}^Ms_k^{-1}\right] \nonumber \\ \exp\left\lbrace-\frac{b}{2}\sum_{k=1}^M\frac{(d_k-f_k)^2}{s_k^2}\right\rbrace.\label{eq|LHpar}
\end{eqnarray}
The resulting posterior probability density will therefore be marginalised over $b$. This procedure is only used for the parameter estimation of model $M_1$. 
The noise scaling parameter is a scale parameter, and a Jeffreys prior will be used:
\begin{equation}
	p(b|M_1) = \frac{1}{b\ln \left( \frac{b_\mathrm{max}}{b_\mathrm{min}}\right)  }.
\end{equation}
The noise scaling parameter for the parameter estimation varies from $b_\mathrm{min}=0.1$ to $b_\mathrm{max}=2$, corresponding to an underestimation of the variance by a factor of 10 and an overestimation of the variance by a factor of 2, respectively. This conservative range can be verified a posteriori, by the probability density marginalised for $b$.

\section{Numerical tests}
\label{sec|test}

In order to demonstrate our Bayesian method, we have simulated some realistic synthetic observations, and analysed them using our procedure.

\subsection{Ideal non detection}

We start with a simple test representing an ideal non-detection. 
The local intensity line profiles are represented by Gaussians, with the width of the point spread function of an instrument with spectral resolution of 4.2\,km\,s$^{-1}$ (R=65000), typical of current high-resolution spectropolarimeters like ESPaDOnS, and an extra turbulent broadening of 5\,km\,s$^{-1}$, typical of the broadening encountered in hot stars. We used a shallow line with a depth of 0.17\,$I_\mathrm{c}$. The rotational velocity field is set to $v\sin i=54$\,km\,s$^{-1}$.
The associated circular polarisation is set to zero (i.e. no magnetic field) and the error bars are set to represent a certain s/n.  
We used two simulated observations corresponding to a s/n of 12\,000 and 24\,000 respectively, values that can typically be achieved for the LSD profiles of hot stars. The effective Land\'e factor and the wavelength of the spectral line are set to 1.2 and 5\,000\,\AA\ respectively.
We performed the Bayesian analysis on a grid extending up to 1\,kG, with a maximum rotation velocity of 700\,km\,s$^{-1}$, leading to a minimum inclination of $8^\circ$. 

\begin{figure}
\includegraphics[width=84mm]{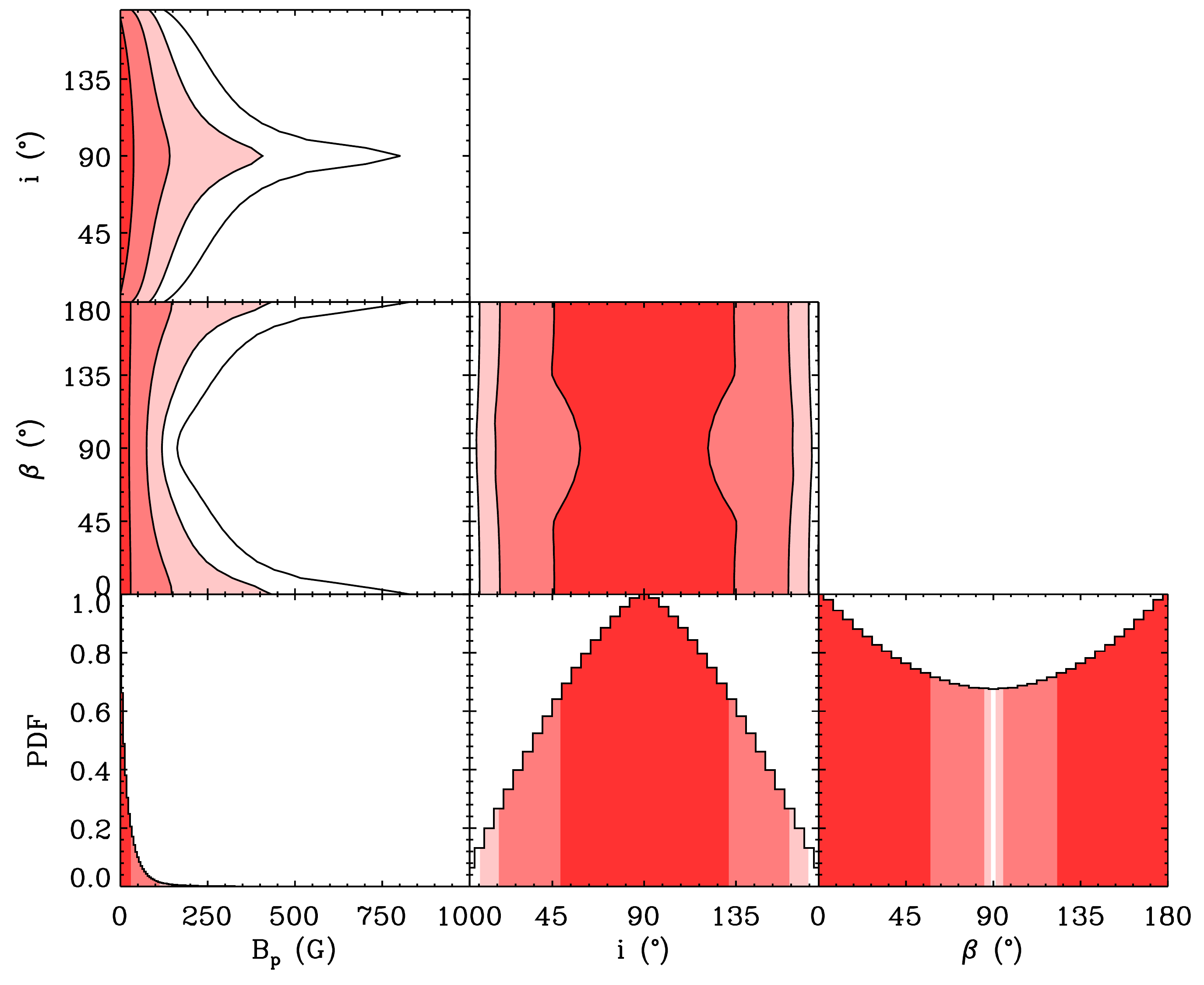}
\caption{\label{fig|mar_tri_nodetect1} Parameter estimation for the ideal non-detection (low s/n). \textit{Bottom:} Posterior probability density functions (PDF) marginalised for the dipole field strength, the rotational axis inclination and the magnetic obliquity, from left to right respectively. The PDFs have been normalised by their maximum. \textit{Top:} 2-D posterior probability density marginalised for the $B_p$-$i$, $B_p$-$\beta$ and $i$-$\beta$ planes. The 68.3\%, 95.4\%, 99.0\% and 99.7\% credible regions are shaded in dark to pale colours respectively. }
\end{figure}

Figure \ref{fig|mar_tri_nodetect1} shows the probability densities resulting from our Bayesian analysis, for the s/n of 12\,000. The three bottom panels show the posterior probability densities marginalised for each parameter -- $B_p$, $i$ and $\beta$, from left to right respectively. The densities have been normalised by their maximum, in order to facilitate the display. The three top panels show the 2D posterior probability densities, marginalised for the $B_p$-$\beta$, $B_p$-$i$ and $i$-$\beta$ planes. 
The contours encircle the regions that contain 68.3\%, 95.4\%, 99.0\% and 99.7\% of the probability. 

As it can be seen from the $B_p$ probability density distribution, the bulk of the probability is concentrated at low $B_p$ values. The amplitude of the Stokes $V$ profiles produced for low $B_p$ values will remain under the noise level, for all possible dipole orientations. As $B_p$ increases, the amplitude of the Stokes $V$ profiles will progressively grow over the noise level, and only a restricted number of possible orientations will produce Stokes $V$ profiles with amplitudes below the noise level, as explained in section \ref{sec|disk}. The probability density marginalised for $B_p$ therefore has a decreasing exponential-like shape, with a tail extending far away from the mode of the distribution. 

The shape of the 2D $B_p$-$i$ and $B_p$-$\beta$ planes gives us insight about the geometries that are more likely to produce a non-detection for a large field strength value. 
When we look at a star rotational equator-on ($i=90^\circ$), the obliquity required to obtain a low amplitude Stokes $V$ profile is around $0^\circ$ or $180^\circ$. Thus, as the field is nearly aligned with the rotational axis, the amplitude of the Stokes $V$ profiles will stay small for the whole phase range. For lower inclination angles, the Stokes $V$ signal will vanish only when the dipole field is seen nearly magnetic equator-on, therefore such an null observation is only possible on a restricted interval of rotation phases (as it was shown in Figure \ref{fig|model_nodetect1_cont2}). 
Consequently, if we assume that a strong field is present but is not producing any Stokes $V$ signal, it is more likely that we are observing a dipole configuration for which the Stokes $V$ profiles always have a small amplitude, rather than a dipole configuration for which the Stokes $V$ signal vanishes for only a small fraction of the rotational period. This explains why the probability density is more extended toward higher $B_p$ values for $i=90^\circ$ and $\beta=0^\circ, 180^\circ$. 

The 2D $i$-$\beta$ plane shows that we have no constraint on the inclination and obliquity angles. The shape of the probability density marginalised for $i$ is in part due to the extra probability at higher $B_p$ for $i\sim90^\circ$, but is mainly driven by the prior probability that favours large inclinations (as given in Eq. (\ref{eq|prior_i})). Following the same reasoning, the aligned dipoles ($\beta\sim0^\circ$ or $180^\circ$) are slightly favoured.

Table \ref{tab|nodetect1} compiles the base ten logarithm of the odds ratio computed for each observation taken individually, and for the two observations combined together. In order to illustrate the effect of the priors, we also computed the odds ratios while using a flat prior for all the parameters. 
For a single low s/n observation, the model $M_0$ for the absence of a magnetic field is preferred by a factor of 2. When doubling the s/n, the odds in favour of $M_0$ only increase to a factor 2.8. When the two observations are combined together, the odds ratio goes up to a factor of 3.   
\begin{table}
	\caption{\label{tab|nodetect1}Base ten logarithm of the odds ratio $\log(M_0/M_1)$ for the ideal non-detections. }
	\begin{center}
	\begin{tabular}{l c c c}
	\hline
	\multicolumn{1}{c}{Test} & $\log(M_0/M_1)$& $\log(M_0/M_1)$& $\log(M_0/M_1)$ \\
	  & low s/n & high s/n & joint obs.	\\
	 \multicolumn{1}{c}{(1)} & (2) & (3) & (4) \\
	\hline
	Flat prior	& 0.976 & 1.267 & 1.355	\\
	Used prior& 0.317	& 0.444 & 0.475 \\
	\hline	
	\end{tabular}
	\end{center}
\end{table}
	
If we had used a flat prior for all parameters, the odds ratios would have been in favour of the model $M_0$ by a factor of roughly 10 and 20, for the low and high s/n respectively. Jeffreys priors are used when we ignore the exact scale of a parameter. This serves to balance the high scales and the lower scales. In a non-detection case, the bulk of the probability is situated at low $B_p$ values. With a Jeffreys prior, the small scales have as much weight as the larger scales, hence the configurations at high $B_p$ values that produce bad fits dominate less the global likelihood. 
The flat prior effectively gives more weight to the larger scales, hence rejecting more strongly the $M_1$ model. We note here that the $M_0$ model is in fact a subset of the $M_1$ model, given that the field strength range extends down to 0\,G. This means that both models can reproduce this simulated dataset perfectly. However, the $M_1$ model is penalised by its complexity, which is mathematically encoded by the priors. Therefore, for a non-detection, the odds ratios will never be very strongly in favour of the $M_0$ model by a large number, as the discrimination comes mainly from ``Occam's razor''. Furthermore, the posterior probability density is sensitive to the exact choice of the prior in this case.
Given the decreasing exponential behaviour of the probability density, we think that the more conservative Jeffreys prior is more suitable, as it is less eager to reject the $M_1$ model. 

Assuming that the dipole model is true, we can perform a parameter estimation to determine which dipole strengths would be admissible by our observations. This can be achieved using the marginalised probability density for $B_p$. Integrating this probability density between two $B_p$ values gives the probability that the true value of the field strength lies between these two values. 
The probability density itself can be used in many applications, such as building a statistical field strength distribution for a stellar population. 
 
In other applications, such as the study of magnetically confined wind shocks, it would be valuable to get an upper field strength limit. When a probability density has a  Gaussian-like shape, it is customary to express our confidence intervals by regions enclosing a certain percentage of the total posterior probability, namely 68.3\%, 95.4\% and 99.7\%. With a Gaussian distribution, the 95.4\% region will be twice as extended as the 63.8\% region, and the 99.7\% will be 3 times as extended as the 68.3\% region, analogous to the 1, 2, 3 $\sigma$ contours in frequentist statistics. 
However, as our probability distribution does not have a Gaussian shape, but is rather shaped like a decreasing exponential, these credible regions will not be regular (i.e. the 99.7\% credible region will reach much farther than 3 times the 68.3\% credible region, taking into account the extended tail of the distribution). We therefore added a credible region enclosing 99.0\% of the probability. 

\begin{table}
	\caption{\label{tab|nodetect1_region} Credible region upper limits for the ideal non-detection.}
\begin{center}
	\begin{tabular}{l c c c c}
	\hline
	\multicolumn{1}{c}{Test} &  68.3\%	& 95.4\%	& 99.0\%	&	99.7\%	\\
	 & (G) & (G) & (G) & (G) \\
	\multicolumn{1}{c}{(1)} & (2) & (3) & (4) & (5) \\
	\hline
	low s/n	& 30 & 112 & 258 & 456 	\\
	high s/n	& 18 &  66  & 157 & 297	 \\
	Joint		& 16 &  51  & 110 & 196 \\
	Joint, flat prior 	& 29	& 138 & 398 & 734 \\
	Joint, no scale noise	& 20 & 67 & 142 & 251 \\
	\hline	
	\end{tabular}
	\end{center}
\end{table}

Table \ref{tab|nodetect1_region} gives the upper limits of the credible regions (the lower limit being 0\,G in each case) for each percentage threshold of the probability. 
For example, the probability that the field strength is lower than 258\,G is 99.0\% for the low s/n case. For the higher s/n case, all the credible regions are narrower. When we combine the two observations together, the 68.3\% and 95.4\% regions are similar to the high s/n case, but the probability density is more peaked (i.e. it has a less extended tail toward the high $B_p$ values), as can be seen from the narrower 99.0\% and 99.7\% credible regions (110\,G and 196\,G respectively). 
When combining two non-detection observations, the probability density of high $B_p$ values narrows around $i=90^\circ$ and $\beta=0^\circ$. The likelihood of the $i=90^\circ$, $\beta=0^\circ$ configurations stays the same, as none of the phases produce any Stokes $V$ signal. However, the likelihood of the other degenerate configurations decreases, as it becomes less likely that we have observed the star both times during the same narrow range of phases that do not produce a Stokes $V$ signal, if the observations were taken at random times. Therefore, the high-$B_p$ tail becomes less important as we combine multiple observations. 

Table \ref{tab|nodetect1_region} also gives the credible regions for the combined observations, while considering a flat prior for each parameter. In that case, the shape of the probability density is more weighted toward the high scales, and the credible regions are more extended toward the high $B_p$ values.

 We also give the credible regions when considering a fixed variance of our data. However, in this particular case, as the observations can be reproduced perfectly by both models (as we set Stokes $V$  strictly to zero), the inclusion of the noise scaling  parameter does not change the probability density much, except for a slight tightening of the credible regions as we allowed for the possibility of variance overestimation.

\subsection{Ideal detections}

In order to show the behaviour of the algorithm in the presence of a magnetic signal, we now add non-zero Stokes $V$ profiles to the simulated observation corresponding to a s/n of 12\,000. Once again, no random noise was added, to illustrate the ideal case. To draw a connection between the credible regions found in the preceding section, we chose field values corresponding roughly to the upper limits of these credible regions: 30\,G, 100\,G, 250\,G and 450\,G. The magnetic configuration was set to $i=90^\circ$ and $\beta=90^\circ$, at the rotational phase when we are looking straight at the magnetic pole ($\varphi=0^\circ$). This corresponds to a profile with a shape like the leftmost profile of Figure \ref{fig|model_nodetect1_cont0}.

Table \ref{tab|nodetect4} gives the odds ratios for the used priors, as well as the ones computed with flat priors for comparison. We also give the credible regions containing 68.3\%, 95.4\%, 99.0\% and 99.7\% of the total probability. The lower limit is 0\,G, unless indicated otherwise. 
\begin{table*}
\begin{minipage}{130mm}
	\caption{\label{tab|nodetect4} Odds ratios ($\log(M_0/M_1)$) for the adopted and flat priors, as well as credible regions (lower limit of 0\,G unless indicated otherwise) for the ideal detections.}
	\begin{tabular}{D{(}{~(}{5,7} c c c c c c}
	\hline
	\multicolumn{1}{c}{Test} & $\log(M_0/M_1)$ & $\log(M_0/M_1)$	& 68.3\% & 95.4\% & 99.0\% & 99.7\%\\
	 & Used prior & Flat prior & (G) & (G) & (G) & (G) \\
	 \multicolumn{1}{c}{(1)} & (2) & (3) & (4) & (5) & (6) & (7) \\
	\hline
	30\,\mathrm{G} (\sim68.3\%)	& 0.303	& 0.939	& 34	& 128	& 290	& 498	\\
	100\,\mathrm{G} (\sim95.4\%)	& 0.0970	& 0.476	&123	& 338	& 625	& 825	\\
	250\,\mathrm{G} (\sim99.0\%)		& -2.98	& -3.22	&196 - 407	& 155 - 763	& 143 - 938	& 141 - 991	\\
	450\,\mathrm{G} (\sim99.7\%)	& -13.4	& -13.9	&410 - 658	& 384 - 928	& 379 - 995	& 363 - 1000	\\
	\hline	
	\end{tabular}
\end{minipage}
\end{table*}

The posterior probability density for the 30\,G observation is indistinguishable from the $V=0$ observation from the previous section. The odds ratio is still marginally in favour of the $M_0$ model, and the credible regions for $B_p$ are similar. 

\begin{figure}
\includegraphics[width=84mm]{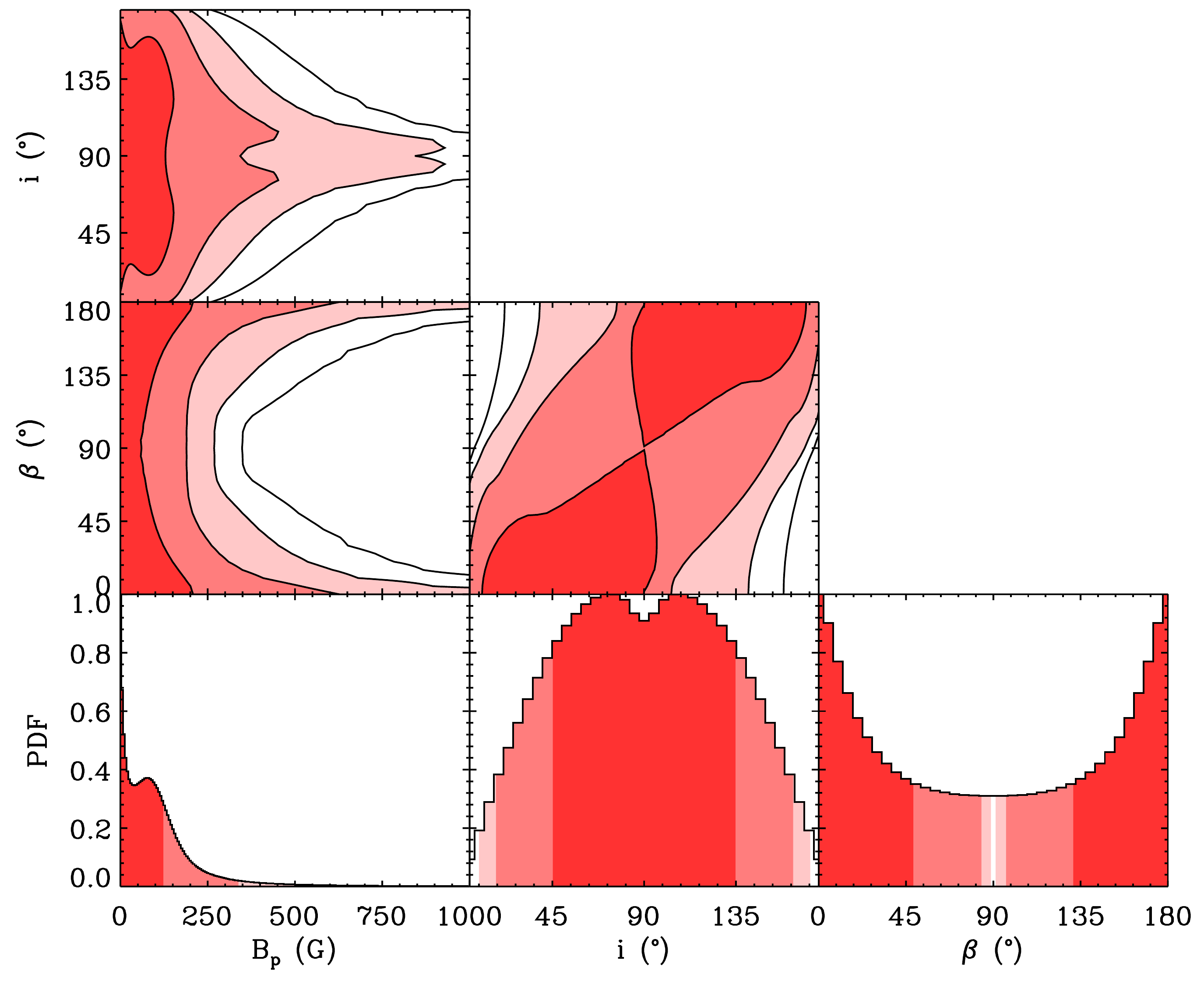}
\caption{\label{fig|mar_tri2_nodetect4} Same as Figure \ref{fig|mar_tri_nodetect1} for an ideal detection with a field strength of 100\,G.}
\end{figure}

At 100\,G, the odds ratio is still in favour of $M_0$, although both models are now nearly as likely ($\log(M_0/M_1=0.097)$). The odds ratio computed with a flat prior still rejects the $M_1$ model by a factor of three. The shape of the probability density, as shown in Figure \ref{fig|mar_tri2_nodetect4}, is also different than the non-detection case, with a secondary peak in the probability density marginalised for $B_p$, although the probability density still peaks at 0\,G. The credible regions for $B_p$ are therefore extended toward higher values, and the 68.3\% credible region upper limit (123\,G) encompasses the real field value.
As it was the case for the ideal non-detection, an aligned magnetic configuration is preferred, as it maximises the chances of observing this particular profile shape. The $i$-$\beta$ 2D plane is different from that of the ideal non-detection. The shape of the magnetic signal adds an extra constraint, by rejecting the configurations for which the positive pole is never located on the visible hemisphere.  

At 250\,G, the odds ratio has switched in favour of the dipole model, by three orders of magnitude ($\log(M_0/M_1)=-2.98$). However, our chosen prior is again more conservative than the flat prior, which favours $M_1$ more strongly ($\log(M_0/M_1)=-3.22$). As a flat prior overweights the larger scales compared to a Jeffreys prior, our used prior needs a more significant likelihood at large $B_p$ values in order to favour the magnetic model. 

\begin{figure}
\includegraphics[width=84mm]{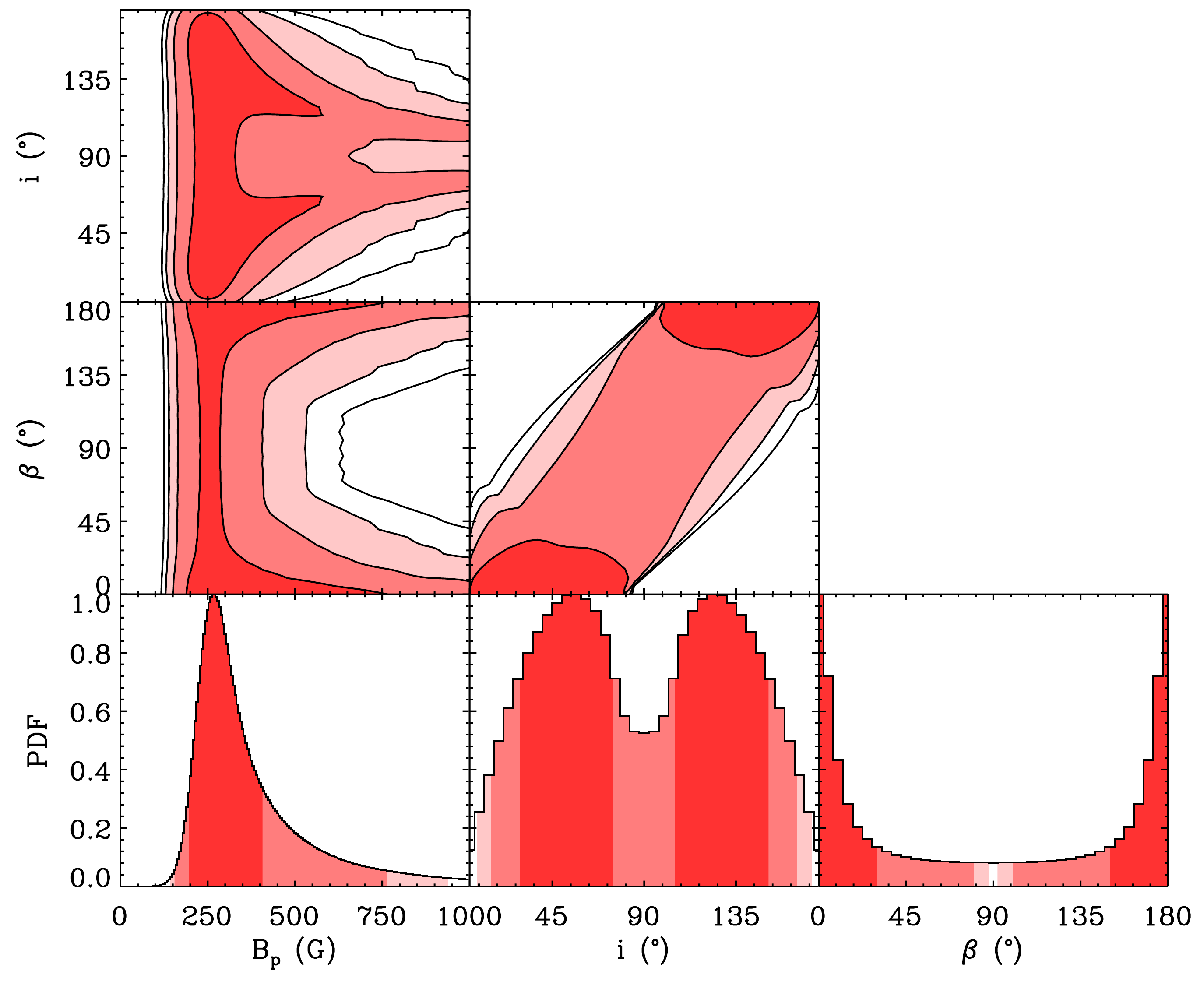}
\caption{\label{fig|mar_tri3_nodetect4}  Same as Figure \ref{fig|mar_tri_nodetect1} for an ideal detection with a field strength of 250\,G.}
\end{figure}

The posterior probability density is now typical of a detectable magnetic field (Figure \ref{fig|mar_tri3_nodetect4}). A sharp cut in the probability density marginalised for $B_p$ and the 2D planes shows a tight constraint on the lower field limit. The $B_p$ distribution peaks around 275\,G and decreases slowly toward higher $B_p$ values. Therefore, the credible regions for $B_p$ are not symmetric with respect to the mode of the distribution. 

The lowest field strength values correspond to configurations for which the positive magnetic pole is located at the stellar disk center. Larger field values are possible when we are looking at the positive pole from an angle. For example, if the field is aligned ($\beta=0^\circ$), the amplitude of the Stokes $V$ profiles will decrease as the inclination decreases, until we reach $i\sim0^\circ$ where the magnetic signal vanishes. 

Because there is only one observation, the aligned configurations are preferred, as shown by the probability density marginalised for $\beta$. 
Obviously, when the inclination is $90^\circ$, an aligned configuration would not provide any Stokes $V$ signal and cannot reproduce the observed profile. Some obliquity would be necessary in order to put the positive pole on the visible hemisphere, which explains the dip at $i=90^\circ$ in the probability density marginalised for $i$ and marginalised for the $B_p$-$i$ plane.
 
When $i\sim0^\circ$ and the field is aligned, we are always looking directly at the positive pole, which makes this configuration quite likely. However, our prior knowledge tells us that a low inclination is less likely than a high inclination. For this reason, the posterior probability density marginalised for $i$ decreases toward $i\sim0^\circ$ and $180^\circ$. 

At the present $\mathcal{B}$ configuration ($i=90^\circ$ and $\beta=90^\circ$) the particular profile shape (anti-symmetrical) only occurs during a certain phase range (see Figure \ref{fig|model_nodetect1_cont0}), which makes such a configuration less likely than the aligned configurations. Therefore, with only one observation, no meaningful constraints can be put on the angles. However, the field strength can be better constrained, and the probability density marginalised for $B_p$ can provide statistical insight into the more likely field strength values.

At 450\,G, the odds ratio favours $M_1$ by more than 13 orders of magnitude. When the $M_0$ model is so strongly rejected, the difference in odds ratios between our chosen prior and the flat prior is less important than when the signal detection was more marginal. This is because the bulk of the likelihood is located at higher $B_p$ values, where the two priors are similar.
The posterior probability density is quite similar in shape to the one with a 250\,G field, with the $B_p$ distributions shifted to higher values, as it can be seen from the credible regions in Table \ref{tab|nodetect4}. 

\subsection{Realistic case}
\label{sec|real}
So far, we have explored the behaviour of the algorithm in cases where the models can reproduce the data perfectly. Obviously, real observations will have some deviations from the predicted values, introduced by the noise. We will now explore how the noise corruption affects our detection capacity. 

\begin{figure}
\includegraphics[width=84mm]{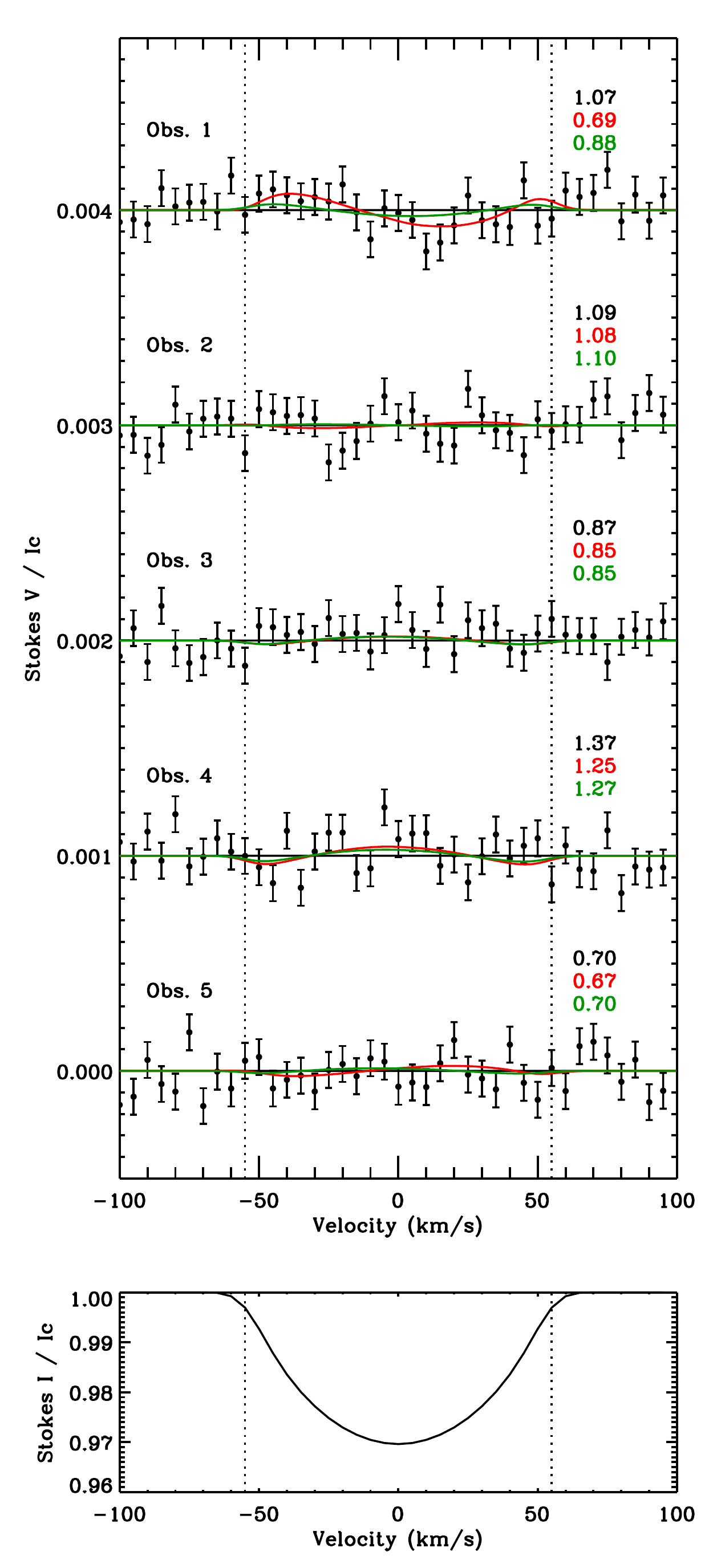}
\caption{\label{fig|profiles_nodetect7} Realistic simulated observations of pure Gaussian noise. The intensity line profile is shown at the bottom, and the five noisy Stokes $V$ profiles are shown at the top. The dotted lines display the range of the fits. The no magnetic field model $M_0$ ($V=0$) is shown by the black lines. The best fits of the oblique dipole model $M_1$ for each observation taken individually (represented by the maximum of the individual likelihoods) are shown in red. The best fit for all the observations taken together by a single $\mathcal{B}$ geometry (the maximum of the joint likelihood) is shown in green. The corresponding reduced $\chi^2$s are given on the right side, with matching colours.}
\end{figure}

\begin{figure}
\includegraphics[width=84mm]{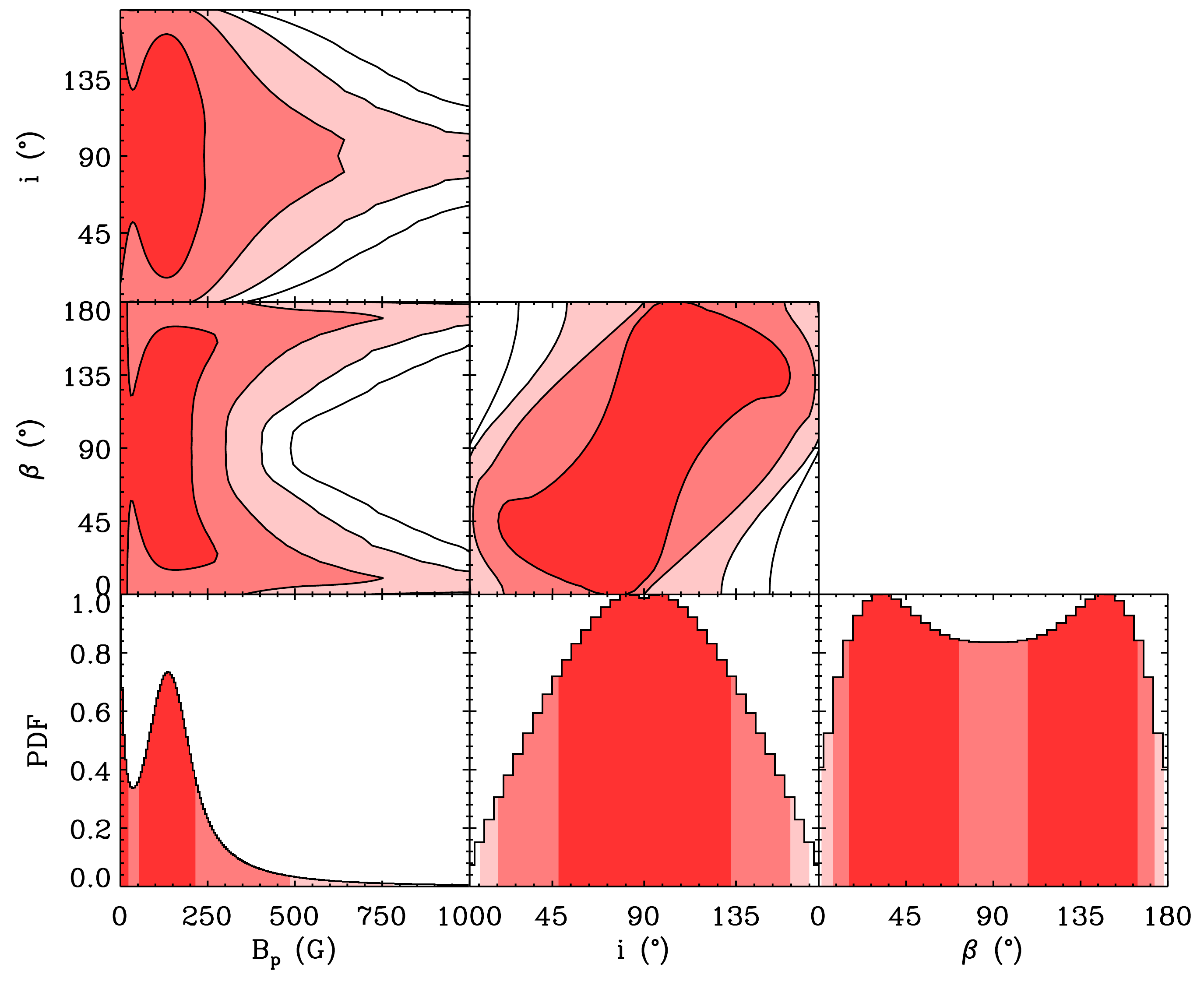}
\caption{\label{fig|mar_tri4_nodetect7} Same as Figure \ref{fig|mar_tri_nodetect1} for Obs. 1 of the realistic simulated observations of pure noise. }
\end{figure}

The top spectrum of Figure \ref{fig|profiles_nodetect7} shows a simulated observation, consisting of only random noise generated from a normal distribution corresponding to a s/n of 12\,000. Figure \ref{fig|mar_tri4_nodetect7} presents the posterior probability density for that specific observation. Notice how the $B_p$ distribution looks like the probability density for the 100\,G observation of the previous section (Figure \ref{fig|mar_tri3_nodetect4})
\footnote{The $\beta$ density probability is different from the 100\,G ideal detection because the simulated noise pattern has a shape similar to a magnetic field whose pole is located at a non-null rotational radial velocity. Such a location requires a dipole that is not aligned.}. 
In fact, the odds ratio is even in  favour of the dipole model ($\log(M_0/M_1)=-0.419$), even though the signal is pure noise. 
This is because the observation is better fitted by a non-null magnetic field. 
The absolute best fit to the data is given by the maximum of the likelihood. We illustrate this fit by the red profile overplotted on the data (Figure \ref{fig|profiles_nodetect7}). 
Therefore, the detection of a real signal embedded in the noise is ambiguous, as the noise could have -- by chance -- the shape of a magnetic profile. An observer would likely not feel confident to report a magnetic detection based on only one observation like this one. 

In order to verify if this ambiguity can be lifted by re-observing the star, we generated 4 additional simulated observations, again with pure normal noise (rest of Figure \ref{fig|profiles_nodetect7}). Each of these observations leads to an odds ratio in favour of $M_0$ by about a factor of two (Table \ref{tab|nodetect6}, column 2). 

The best fit achievable for each observation taken individually is again overplotted in red. However, nothing restricts the magnetic configuration $\mathcal{B}$ to be the same for each observation. The best fit produced by a single $\mathcal{B}$ configuration is given by the maximum of the joint likelihood, illustrated in green.
Although the data can be reproduced by a non null magnetic field, the odds ratio of the combined observations is in favour of the $M_0$ model ($\log(M_0/M_1)=0.295$). The slight improvement of the fit produced by the oblique dipole model (see the reduced $\chi^2$ indicated in Figure \ref{fig|profiles_nodetect7}), is not enough to justify the use of a more complex model. 
\begin{table}
	\caption{\label{tab|nodetect6} Odds ratios ($\log(M_0/M_1)$) for the realistic simulated observations for the pure noise case and the $B_p=125$\,G case. The rotational phases, chosen randomly, are also given. }
\begin{center}
	\begin{tabular}{l c c c}
	\hline
	\multicolumn{1}{c}{Test} 				& $\log(M_0/M_1)$ &$\varphi_n$& $\log(M_0/M_1)$ \\
					& Noise only	&					&$B_p=125$\,G		\\
	\multicolumn{1}{c}{(1)}				&(2)			&(3)					&(4)\\				
	\hline
	Obs. 1			& -0.419	&0.76& -4.04	\\
	Obs. 2			& 0.306	&0.43& -0.412	\\
	Obs. 3			& 0.295	&0.40& -0.484	\\
	Obs. 4			& 0.178	&0.20& -2.26	\\
	Obs. 5			& 0.281	&0.60& -0.388	\\
	Combination		& 0.295	&	& -9.05	\\
	\hline	
	\end{tabular}
	\end{center}
\end{table}

\begin{figure}
\includegraphics[width=84mm]{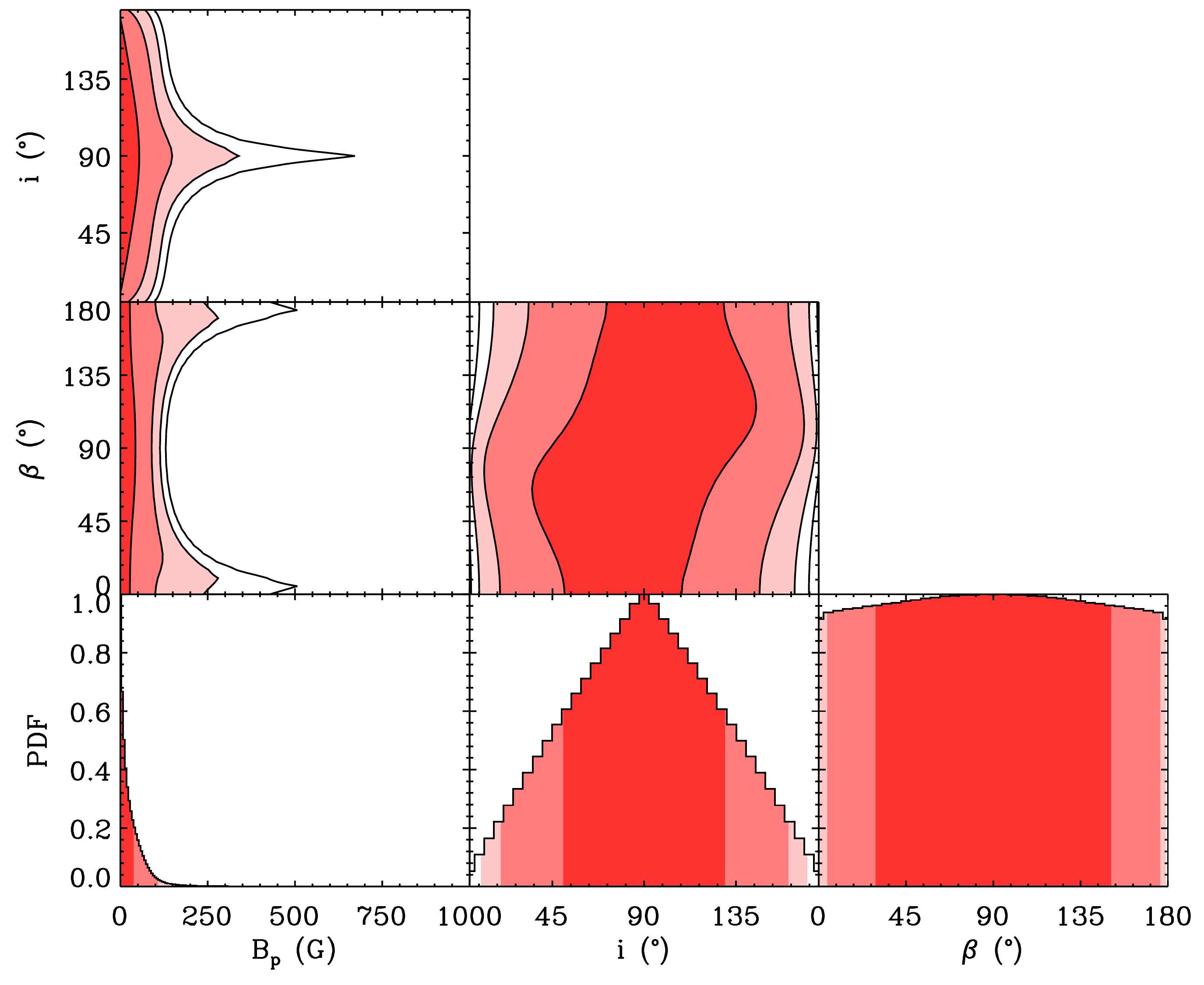}
\caption{\label{fig|mar_tri_nodetect7}  Same as Figure \ref{fig|mar_tri_nodetect1} for the combined realistic simulated observations of pure noise.}
\end{figure}

The posterior probability density (Figure \ref{fig|mar_tri_nodetect7}) is similar in shape to the perfect non-detection that was shown in Figure \ref{fig|mar_tri_nodetect1}. The high-$B_p$ tail is less extended that in the case of a single observation. This can by seen more easily by comparing the 2D probability densities for the $B_p$-$\beta$ and $B_p$-$i$ planes of Figures \ref{fig|mar_tri_nodetect7} and \ref{fig|mar_tri_nodetect1}. 
As explained in the previous sections, combining multiple observations has this effect because, in the case of non-detections, the high-$B_p$ inclined dipole configurations become less likely, as this would mean that all the observations were taken during a narrow range of phase. 
Obviously, the possibility that the observations were taken at the same rotational phase can generally be assessed by the time span of the observations and the possible range of rotational periods. In the case where it is known that the rotational period is quite long compared to the observation time span, more information is available than is assumed in this algorithm (that the phase of each observation can assume any value). In that case, it would be wiser to combine all the observations together and treat them as a single observation in order to get a more meaningful probability density. 

Table \ref{tab|nodetect7} compiles upper limits to the 68.3\%, 95.4\%, 99.0\% and 99.7\% credible regions, extracted from the posterior probability density marginalised for $B_p$. Although we combined five observations, the 68.3\% credible region for $B_p$ extends to 38\,G, which is higher than the value obtained for one perfect observation of low s/n (30\,G), because of the deviations introduced by the noise. However, as mentioned above, the $B_p$ distribution of the combined observations does not extend as far than as that from a single perfect observation, as illustrated by the 99.7\% credible region that extends only to 370\,G, compared to 456\,G for the perfect observation. 
We also compiled the credible regions we would obtain without the use of the noise scaling parameter. As expected, the deviations from the model are consistent with our estimation of the variance (the error bars) and the credible regions are nearly the same in both cases. 
\begin{table}
	\caption{\label{tab|nodetect7} Credible region upper limits for the combined realistic simulated observations of pure noise. }
\begin{center}
	\begin{tabular}{l c c c c }
	\hline
	\multicolumn{1}{c}{Test} &  68.3\% & 95.4\% & 99.0\% & 99.7\%\\
		& (G) & (G) & (G) & (G) \\
	\multicolumn{1}{c}{(1)} & (2) & (3) & (4) & (5) \\	
	\hline
	With noise scaling		& 38	& 108	& 214	& 370	\\
	Without noise scaling & 38 & 107 & 213 & 368 \\
	\hline	
	\end{tabular}
	\end{center}
\end{table}

\begin{figure}
\includegraphics[width=84mm]{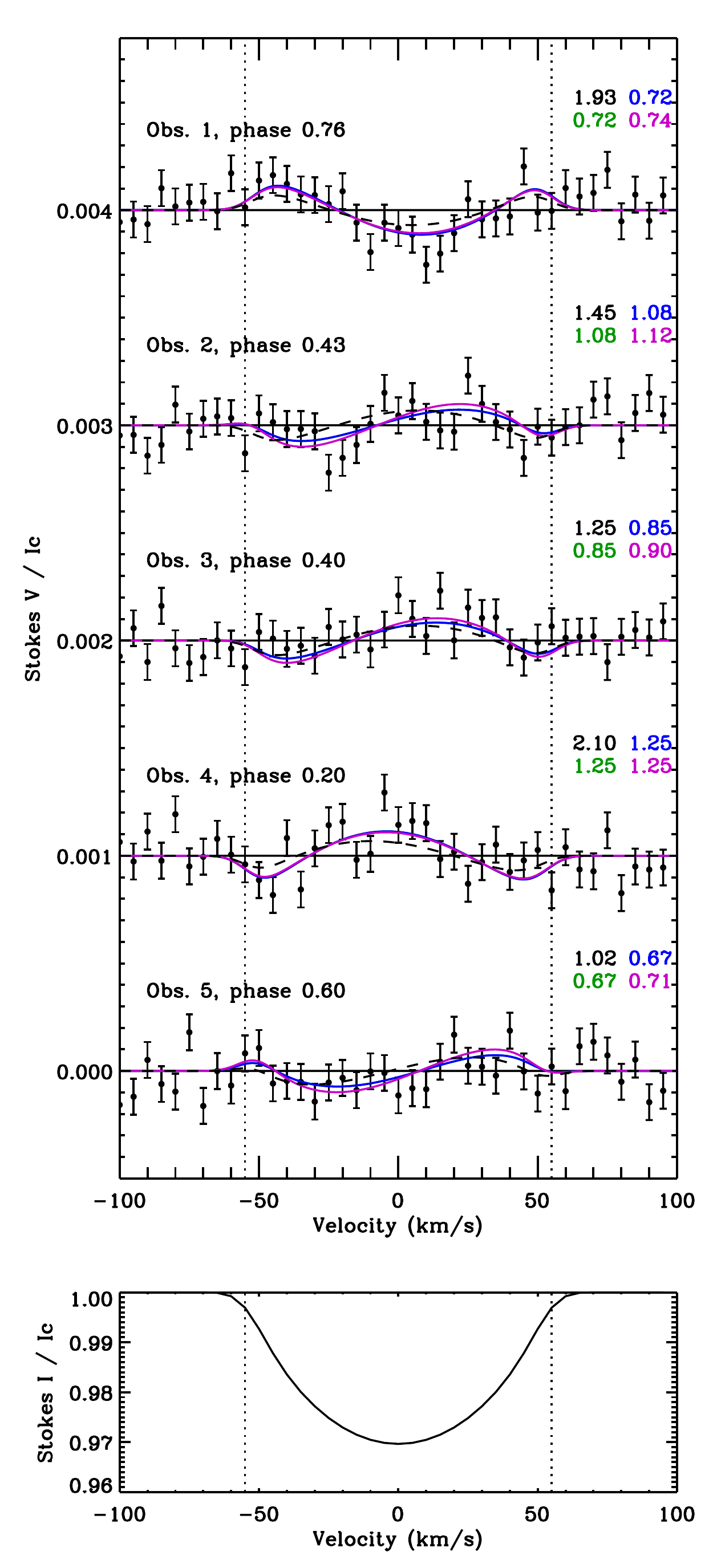}
\caption{\label{fig|profiles_nodetect6} Realistic simulated observations of noise with a Stokes $V$ signal corresponding to $B_p=125$\,G ($i=\beta=90^\circ$). The intensity line profile is shown at the bottom, and the five noisy Stokes $V$ profiles are shown at the top. The dotted lines display the range of the fits. 
The underlying real Stokes $V$ signals are shown with black dashed curves.
The null magnetic field model $M_0$ ($V=0$) is shown by the black lines. The best fit by a single $\mathcal{B}$ geometry for all the observations taken together (the maximum of the joint likelihood) is shown in green. The fit produced by the MAP $\mathcal{B}$ parameters is shown in blue, and that produced by the modes of the posterior probability density marginalised for each parameter is shown in magenta. 
The corresponding reduced $\chi^2$s are given on the right side, with matching colours.
}
\end{figure}

We have therefore demonstrated that a suspicious signal that has a shape similar to that of a magnetic signal can be verified by the acquisition of a small number of additional observations. We now demonstrate that the same is true for a real signal that is sufficiently embedded in the noise to render the odds ratio of a single observation ambiguous. 
Given that in the ideal detection case we were able to detect a field with a strength near the 95.4\% upper limit of the ideal non-detection, we chose a field strength close to the upper limit of the 95.4\% credible region of the previous example. The $\mathcal{B}$ configuration is given by $B_p=125$\,G, $i=90^\circ$ and $\beta=90^\circ$. Five phases were randomly generated ($\varphi= 0.76,\, 0.43,\, 0.40,\, 0.20,\, 0.60$), and the corresponding Stokes $V$ profiles were added to the previous simulated dataset of pure noise. The resulting simulated observations are shown in Figure \ref{fig|profiles_nodetect6}. The underlying real Stokes $V$ profile is shown as the black dashed curve for each observation. 

Table \ref{tab|nodetect6} gives the odd ratio for each observation (column 4), all of which favour the dipole model $M_1$, although observations 2, 3 and 5 less strongly than observations 1 and 4. In fact, the odds ratios of the formers are similar to the odds ratio of the first observation of the pure noise test and on their own, each of these observations would result in an ambiguous signal detection. However, combining all the observations together, the odds ratio is now strongly in favour of the dipole model by 9 orders of magnitude ($\log(M_0/M_1)=-9.05$). 

\begin{figure}
\includegraphics[width=84mm]{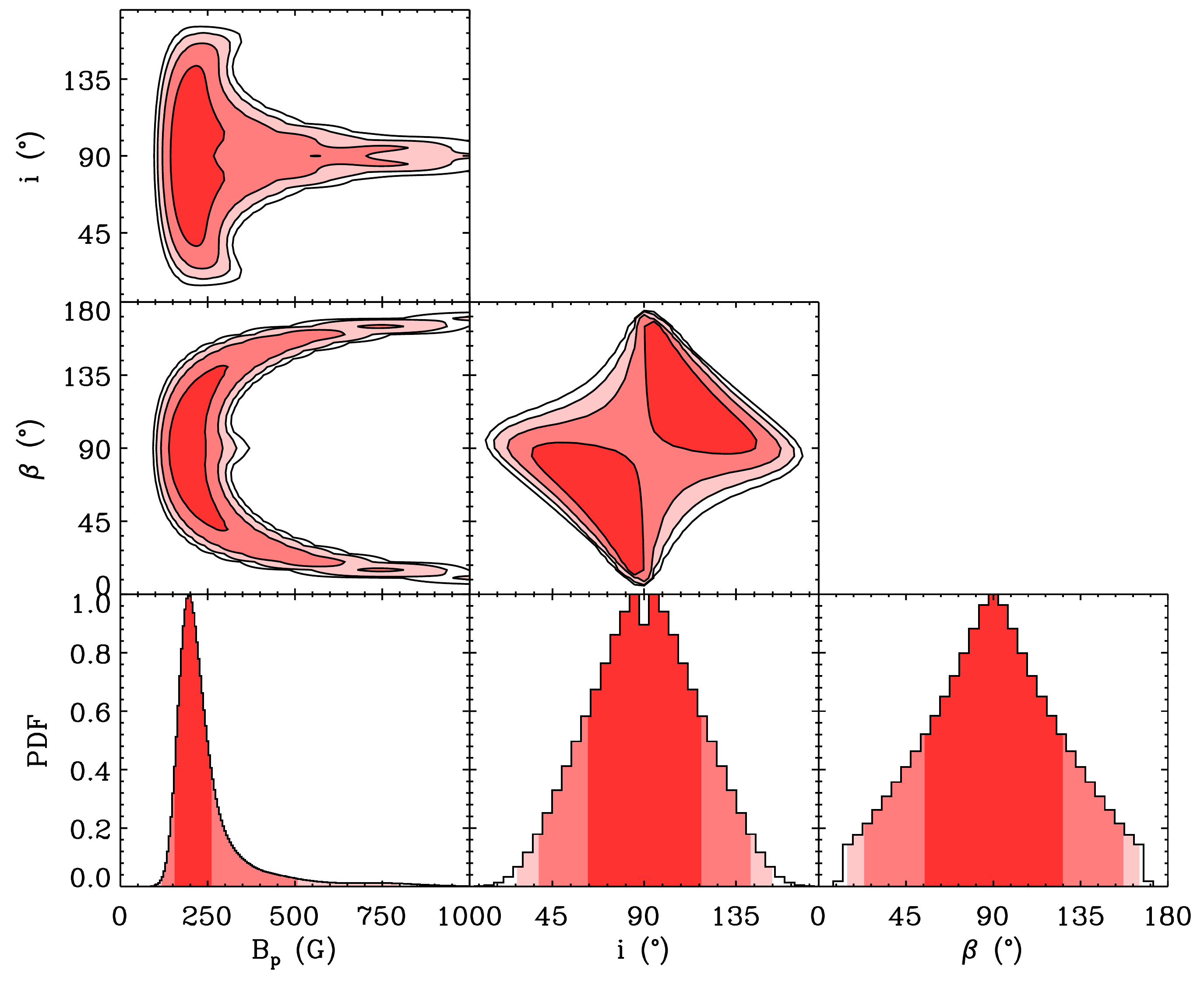}
\caption{\label{fig|mar_tri_nodetect6} Same as Figure \ref{fig|mar_tri_nodetect1} for the combined realistic simulated observations of noise plus a signal corresponding to $B_p=125$\,G.}
\end{figure}

Figure \ref{fig|mar_tri_nodetect6} shows the posterior probability densities for the combined observations. There is a sharp lower limit on the possible magnetic strengths, illustrated by the probability density marginalised for $B_p$, and for the 2D $B_p$-$i$ and $B_p$-$\beta$ planes. Given that most of the likelihood is situated at non-null field strengths, the improvement of the fit to the data justifies the more complex model, as shown by the reduced $\chi^2$ on Figure \ref{fig|profiles_nodetect6} for the $M_0$ model fits (black lines) and the fit produced by maximum of the joint likelihood for $M_1$ (green curves). 

\begin{table*}
\begin{minipage}{130mm}
	\caption{\label{tab|nodetect6_region} Estimation of the $\mathcal{B}$ parameters for the realistic simulated observations of noise plus a signal corresponding to $B_p=125$\,G. The first three columns give the maximum of the joint posteriori probability density as well as the modes and the medians of the joint posteriori probability marginalised for each parameter. The last four columns give the credible regions of each parameter. }
\begin{center}
	\begin{tabular}{l c c c c c c c}
	\hline
	\multicolumn{1}{c}{Parameter} & MAP & Mode &  Median & {68.3\%} & {95.4\%} & {99.0\%} & {99.7\%}\\ 
	\multicolumn{1}{c}{(1)} & (2) & (3) & (4) & (5) & {(6)} & {(7)} & {(8)}\\
	\hline
	$B_p$ (G)  & 210 & 195 & 220 & 155 - 261 & 118 - 508 & 109 - 793 & 102 - 892 \\
	
	$i$ ($^\circ$)  &61 & 85 & 71 & 62 - 118 & 38 - 142 & 27 - 152 & 21 - 158 \\
			     & 119 & 95 & 109 & & & & \\
	$\beta$ ($^\circ$) & 70 & 90 & 66 & 54 - 125 & 23 - 157 & 14 - 165 & 12 - 168\\
				   & 110 &      & 113 &   &  &   & \\
	\hline	
	\end{tabular}\end{center}
\end{minipage}
\end{table*}

Different choices are possible in order to express the derived values for the model parameters. One can choose for example the maximum of the joint posterior probability density (MAP), or the mode of the marginalised probability density for each parameter. Note however that if the probability distribution is complex, the parameters given by the MAP do not necessarily correspond to the mode of the marginalised probability densities. Usually, the MAP will produce the best fit to the data given that the a-priori information does not exclude any interesting parts of the parameter space, but does not necessarily represent the bulk of the probability. The mode of each parameter represents well the bulk of the probability, but does not necessarily give an excellent fit to the data. Using the median of the marginalised probability densities is usually a good compromise \citep{2005blda.book.....G}. 

In Table \ref{tab|nodetect6_region}, we list the MAP, the mode and the median for each $\mathcal{B}$ parameter, which in this case are quite similar. The MAP, the modes and the medians all yield similarly good fits to the data (the MAP and the modes fits are shown in Figure \ref{fig|profiles_nodetect6} in blue and magenta respectively). 

The range of the credible regions of the probability density marginalised for each parameter are compiled in Table \ref{tab|nodetect6_region}. 
Using the median with the 68.3\% credible region, we would infer a dipole strength $B_p=220\stackrel{+41}{_{-65}}$\,G, which is slightly higher than the input dipole field strength of 125\,G. We verified that this difference was due to the particular noise pattern used in the data, as additional noise simulations allowed us to recover the input value within the 68.3\% region. The real values of the $i$ and $\beta$ angles are recovered by the 68.3\% credible regions, although the constraints are poor, as expected.

\subsection{Comparison with traditional diagnostics}

With low-resolution instruments (such as FORS1 and FORS2 at the Very Large Telescope), one is generally only sensitive to the global longitudinal component of the magnetic field, as the rotationally-broadened spectral lines are not resolved. This global longitudinal field value is extracted from the spectrum using the ``slope'' method as described in \citet{2002A&A...389..191B,2006A&A...450..777B}.  
The presence of a magnetic field in a single observation is therefore diagnosed by the significance of the global longitudinal field measurement compared to its error bar.

In the case of high-resolution spectropolarimetry (e.g. ESPaDOnS at Canada-France-Hawaii Telescope, Narval at T\'elescope Bernard-Lyot or HARPSPol at ESO La Silla 3.6\,m Telescope), the rotationally-broadened spectral lines are resolved and the field is generally diagnosed by the deviation of the circular polarisation profile with respect to $V=0$. The detection can be quantified by the probability that such a deviation from $V=0$ is produced by random noise (the false alarm probability or FAP). The detection probability $P$ can then be expressed as $P=1-FAP$.  
Following \citet{1997MNRAS.291..658D}, a field is generally considered detected if the FAP is less than $10^{-5}$ ($P>99.999\%$), and marginally detected when FAP is less than $10^{-3}$ ($P>99.9\%$). 
With high-resolution spectropolarimetry, it is also possible to integrate the signal over the line profile in order to recover a value equivalent to the global longitudinal field obtained from low-resolution instruments \citep{1997MNRAS.291..658D,2000MNRAS.313..851W}. 
Although it is possible to detect a magnetic signal even when the global longitudinal field is null, this is a useful quantity for producing longitudinal field curves and for comparing with low-resolution data. 
\begin{table}
	\caption{\label{tab_bl} Traditional field diagnostics applied to our numerical tests. The global longitudinal field was integrated from $-55$\,km\,s$^{-1}$ to $+55$\,km\,s$^{-1}$. }
	\begin{center}
	\begin{tabular}{l D{*}{\,\pm\,}{3,3} c c  }
	\hline
	\multicolumn{1}{c}{Test} & \multicolumn{1}{c}{$B_l$} & $|B_l/\sigma|$ & $P(V)$ \\
		& \multicolumn{1}{c}{(G)} & & (\%) \\
	\multicolumn{1}{c}{(1)} & \multicolumn{1}{c}{(2)} & (3) & (4) \\ \hline
	\multicolumn{4}{c}{Ideal detections} \\
	~\,30\,G	& 11*33 & 0.34 & 0 \\
	100\,G & 37*33 & 1.12 & 0 \\
	250\,G & 93*33 & 2.76 & 52.0 \\
	450\,G & 168*33 & 5.03 & 99.99997 \\
	\hline
	\multicolumn{4}{c}{Noise only} \\
	Obs. 1 & 72*33 & 2.16 & 60.8 \\
	Obs. 2 & 12*33 & 0.38 & 64.1 \\
	Obs. 3 & 2*33   & 0.08 & 16.7 \\
	Obs. 4 & -16*33 & 0.48 & 87.9 \\
	Obs. 5 & 10*33 & 0.32 & 13.6 \\
	\hline
	\multicolumn{4}{c}{$B_p=125$\,G} \\
	Obs. 1 & 74*33 & 2.22 & 99.5 \\
	Obs. 2 &	 -28*33 & 0.87 &91.9 \\
	Obs. 3 & -35*33 & 1.07 & 64.8 \\
	Obs. 4 &	 0*33 & 0.01 & 99.8 \\
	Obs. 5 & -27*33 & 0.82 & 54.7 \\
\hline
	\end{tabular}
	\end{center}
\end{table}

We applied these traditional diagnostics to our simulated datasets and we report the results in Table \ref{tab_bl}. In the case of the ideal detections, both the longitudinal field (columns 2 and 3) and the detection probability (column 4) yield a detection when the field strength reaches 450\,G. However, the odds ratios are already in favour of the magnetic model at 250\,G. The detection probability only looks at the total deviation, whereas the odds ratio looks for a shape similar to that of the theoretical model. 

For the realistic case consisting only of noise, all five observations have longitudinal field measurements below $3\sigma$ significance (see column 3). Moreover, no signal is detected in the Stokes $V$ profiles, as shown by the detection probabilities. The same is also true for the realistic case consisting of noise plus a signal for $B_p=125$\,G, although the detection probability nearly reaches a marginal detection for Obs. 1 and 4. For these observations, the odds ratios were in favour of the magnetic model by 2 and 4 orders of magnitude respectively. For the remaining observations, the odds ratios were also in favour of the magnetic model, although at a lower significance. 
Therefore, if we had observed any of the simulated observations with $B_p=125\,G$, the traditional diagnostics would not have diagnosed the presence of a field, but the odds ratio would have indicated the possible magnetic signal. A few additional observations are enough to distinguish between real noise and a buried signal consistent with an oblique dipole field, as demonstrated by the example here. Therefore, Bayesian odds ratios provide a quantitative indication of stars worth re-observing in magnetic surveys.

\section{Application to real data}
\label{sec|lpori}
In order to present the application of this method to real data, we will use high-resolution spectropolarimetric observations of the magnetic B-type star LP\,Ori (=Par\,1772, HD\,36982). 

LP\,Ori is often considered a chemically peculiar star because it was first classified as a B1.5p star by \citet{1952ApJ...116..251S}. LP\,Ori's status as a He-strong or He-weak star is still uncertain. An inspection of our seven spectra (described below) does not reveal any significant variation in the He-line strength that would indicate a He-strong or He-weak star. Furthermore, comparing our spectra with the BSTAR grid of synthetic spectra from non-LTE \textsc{tlusty} models \citep{2007ApJS..169...83L}, we find a reasonable agreement with a temperature of 20\,kK, a $\log g$ of 4.0 and a $v\sin i$ of 80\,km\,s$^{-1}$.
LP\,Ori is also a candidate Herbig Ae/Be star, because of its far-infrared excess, although no emission is present in the visible spectrum. \citet{2002MNRAS.334..419M} suggested that LP\,Ori is an object transiting from the pre-main sequence to the main sequence. 
LP\,Ori seems to be a single star. No radial velocity variation was found by \citet{1991ApJ...367..155A} nor was any speckle companion with a K-band ratio of less than 0.04 for a separation of 150\,mas or more \citep{1999NewA....4..531P}. 

The magnetic field of LP\,Ori was first reported by \citet{2008MNRAS.387L..23P}. In that paper, they used three Stokes $V$ observations obtained with the ESPaDOnS and Narval spectropolarimeters to detect the magnetic field.

ESPaDOnS and Narval are twin high-resolution spectropolarimetric instruments located at Canada-France-Hawaii Telescope and T\'elescope Bernard-Lyot respectively. 
A polarisation measurement consists of a set of 4 sub-exposures taken with different polarimeter configurations. From this measurement set, the circular polarisation Stokes $V$ spectrum is extracted, as well as a diagnostic null polarisation spectrum (labeled $N$) by combining the sub-exposures in such a way that the astronomical object's polarisation should cancel out. ESPaDOnS frames were processed using the \textsc{Upena} pipeline provided by CFHT. The Narval frames were processed by the TBL archive pipeline. Both pipelines use the reduction package \textsc{Libre-ESPRIT} \citep{1997MNRAS.291..658D}. The spectral range of both instruments covers the 370\,nm to 1050\,nm wavelength band, with a resolution $R\sim65\,000$.

\citet{2008MNRAS.387L..23P} used the Bayesian method described here to estimate the dipole strength of LP\,Ori, obtaining $1150\stackrel{+320}{_{-200}}$\,G. For the present analysis, we have obtained one additional ESPaDOnS observation (within the context of the MiMeS CFHT Large Program) and an additional 3 Narval observations. The observation log, which includes the new observations as well as those analysed by Petit et al. (2008), is given in Table \ref{tab|lplog}\,\footnote{During the analysis of the complete set of observations of LP\,Ori for this paper, it was discovered that the sign of the Stokes $V$ profiles corresponding to two spectra employed by Petit et al. (2008) was inverted (the Jan. 2006 ESPaDOnS spectrum and the Narval spectrum). During the present analysis we have corrected the sign of these spectra and verified the sign of all others. We have verified that the results of Petit et al. (2008) are not substantially modified by this change.}. The sample of observations has a large range of s/n, and is therefore suitable for testing the behaviour of the Bayesian method on real data.

\begin{table*}
	\caption{\label{tab|lplog} Log of observations for LP\,Ori. The exposure time is given as the total of the sub-exposures. The signal-to-noise ratio is the mean, per 1.8\,km\,s$^{-1}$ spectral pixel, between 500\,nm and 600\,nm. The last six columns give the global longitudinal field, the significance of the longitudinal field measurement and the signal detection probability for both Stokes $V$ and the null $N$ profiles. }
	\begin{tabular}{l l c c c D{*}{\,\pm\,}{4,4} c c  D{*}{\,\pm\,}{4,4} c c }
	\hline
	\multicolumn{1}{c}{Date} & \multicolumn{1}{c}{Instr.} & HJD & t$_\mathrm{exp}$ & s/n & \multicolumn{1}{c}{$B_l$ $V$} & $|B_l/\sigma|$ $V$& P $V$ & \multicolumn{1}{c}{$B_l$ $N$} & $|B_l/\sigma|$ $N$ & P $N$ \\ 
		&		& (+2\,450\,000) & (s) & & \multicolumn{1}{c}{(G)} & &(\%) &  \multicolumn{1}{c}{(G)} & &(\%)  \\	\multicolumn{1}{c}{(1)} &\multicolumn{1}{c}{(2)}&(3)&(4)&(5)&\multicolumn{1}{c}{(6)}&(7)&(8)&\multicolumn{1}{c}{(9)}&(9)&(11)\\ \hline

	2007-11-08 & Nar. & 4413.54984 & 4000 & 73 & 300*434 	&0.69 & 60.0 & 441*436 	& 1.01 &31.9 \\
	2007-11-09 & Nar. & 4414.54299 & 6000 & 170 & 93*174 	&0.53 & 99.8 & -293*173 	& 1.69 &66.2 \\
	2007-11-10 &  Nar. & 4415.50631 & 8800 & 232 & 199*120 	&1.66 & 87.1 & -3*120		& 0.03 &96.7 \\
	2007-11-11 & Nar. & 4416.54958 & 6000 & 244 & 354*113 	&3.13 & \textbf{99.998}  & -18*112  	& 0.17 &9.4 \\
	2006-01-11 & ESP. & 3747.79588 &9600 &  312	& 314*91 	& 3.45 & \textbf{100}  & -244*91 	& 2.67 & 64.0  \\
	2007-03-06 & ESP. & 4166.76514 & 9600 & 510	& 98*50 		& 1.93 & \textbf{100}	 &  -1*51		& 0.02 &66.4 \\
	2010-02-23 & ESP. & 5250.77084 & 6400 & 426 & 347*61 	&5.62 & \textbf{100} 	 & 97*62 		& 1.57 &97.7  \\ \hline
	\end{tabular}
\end{table*}

\begin{figure}
	\includegraphics[width=84mm]{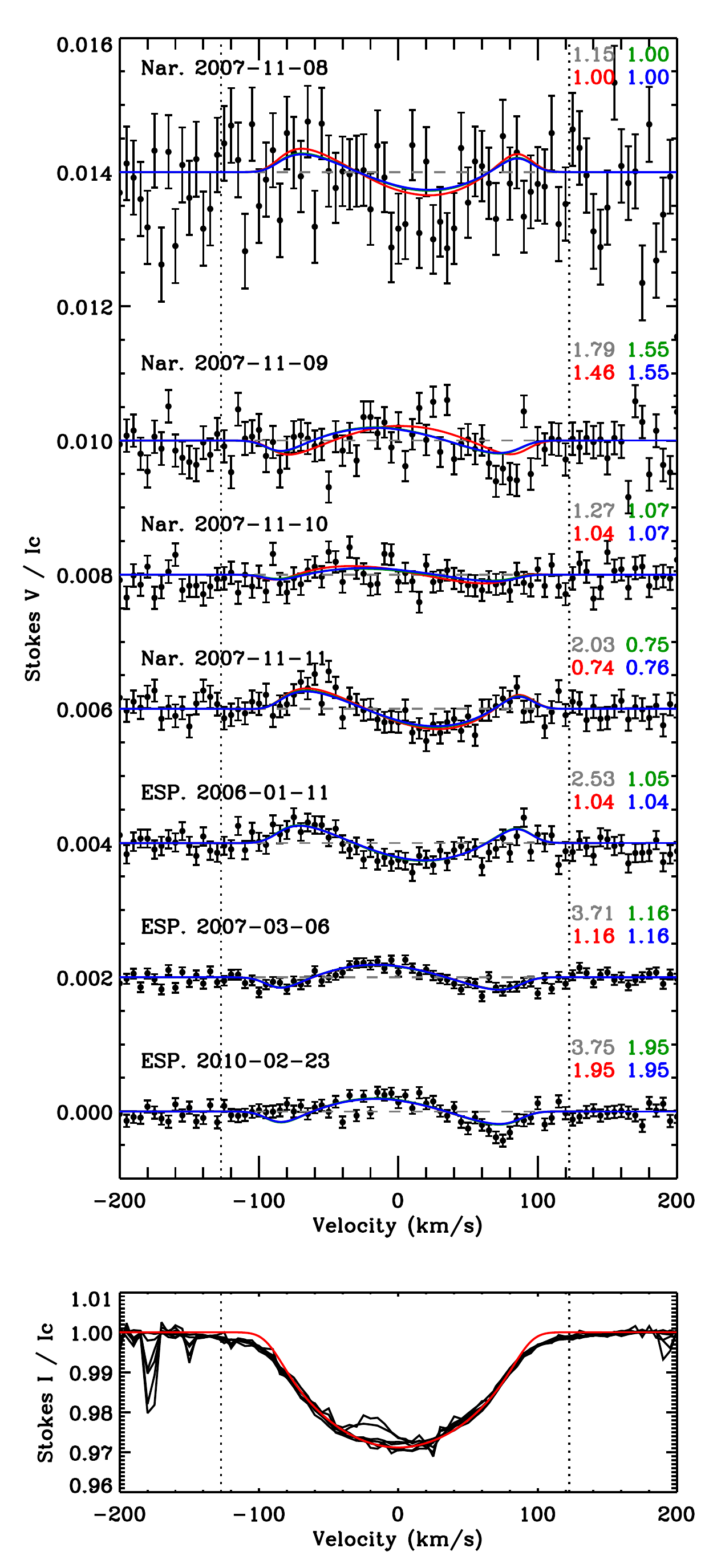}
	\caption{\label{fig|lpV} \textit{Bottom:} LSD line profiles of LP\,Ori (black). The synthetic line profile used for the Bayesian analysis is shown in red. The dashed vertical lines display the range of the fits.  \textit{Top:} LSD Stokes $V$ profiles of LP\,Ori. The grey curves correspond to the $M_0$ model ($V=0$). The red curves correspond to the best fit for each observation taken individually (represented by the maximum of the individual likelihood). The green curves represent the best fit for all the observations taken together by a single $\mathcal{B}$ geometry (the maximum of the joint likelihood). The blue curves correspond to the maximum of the posterior probability density (MAP). The corresponding reduced $\chi^2$s are given on the right side, with corresponding colours. }
\end{figure}

\begin{figure}
	\includegraphics[width=84mm]{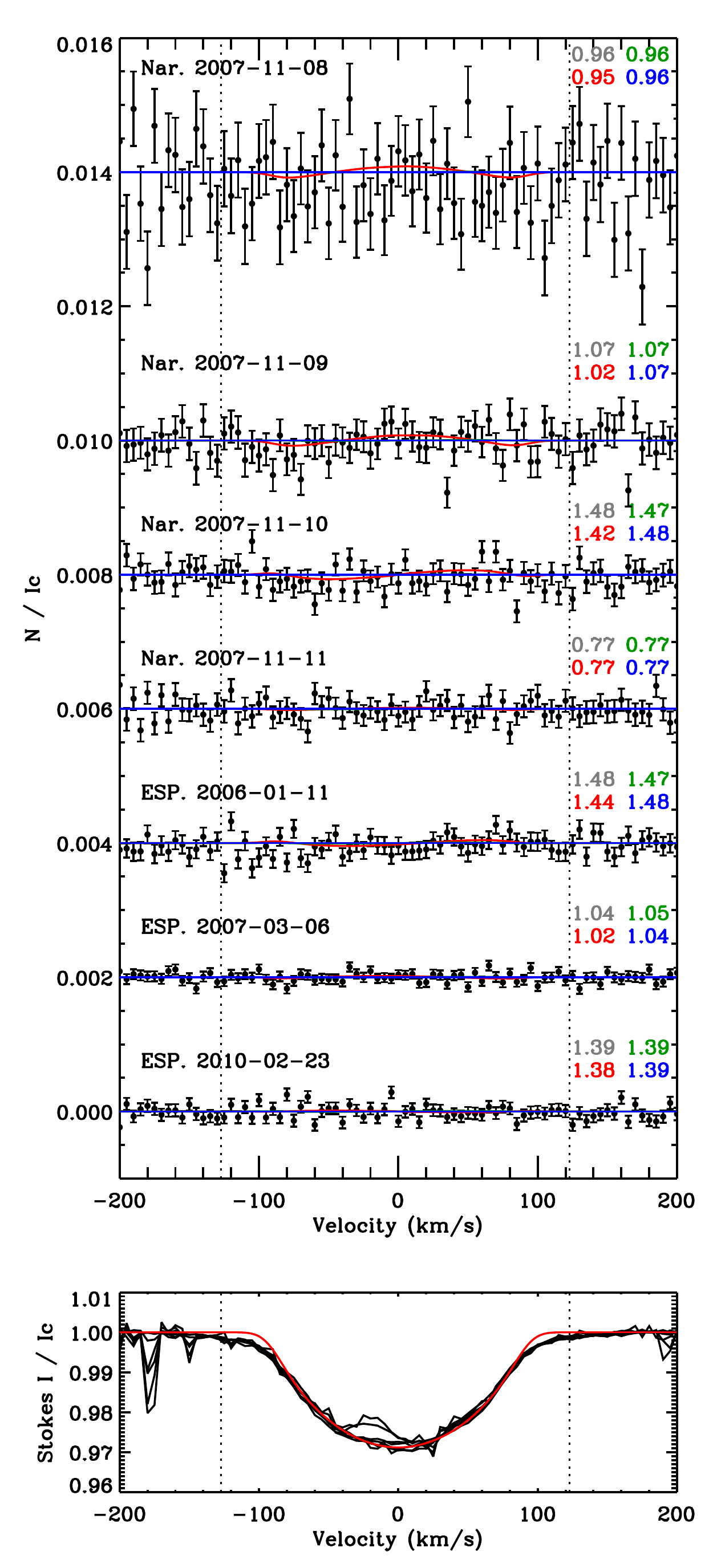}
	\caption{\label{fig|lpN} Same as Figure \ref{fig|lpV} for LP\,Ori null $N$ profiles.  }
\end{figure}

As is customary, we applied the LSD procedure to our observations. We used the i\textsc{LSD} code described by \citet{2010A&A...524A...5K}. We chose spectral lines from a Vienna Atomic Line Database \citep[VALD;][]{2000BaltA...9..590K} list corresponding to an atmosphere model with $T=20$\,kK and $\log g=4.0$. From that list, we chose metallic lines and weak He lines that were unblended with strong Balmer lines and uncontaminated by telluric lines or strong nebular emission. The line depths have been slightly adjusted to match the spectra. The final line list is shown in Table \ref{tab|mask}. The intensity ($d$) and polarisation ($dg\lambda$) weights of each line were normalised by 0.2 and 120 respectively. 
Therefore, the displayed $y$-axis scale of the resulting Stokes $V$ and diagnostic null LSD profiles (Figures \ref{fig|lpV} and \ref{fig|lpN}) corresponds to a line with a $d=0.2$ and $g\lambda=600$. 

The two sharp lines in the blue continuum of the Stokes $I$ LSD profiles are residuals from telluric lines adjacent to the He\textsc{I}\,$\lambda7281$ line. The emission bump present in the LSD profiles of the two first Narval observations are residuals from He emission, most likely of nebular origin because their centroid velocities and scaling match the nebular Balmer line emission. 

We computed the traditional diagnostics -- the detection probability of the individual profiles and the global longitudinal field -- by integrating $\pm115$\,km\,s$^{-1}$ around the line centre, for both the Stokes $V$ and $N$ profiles. The results are displayed in Table \ref{tab|lplog}. 
Definite detections (column 8) are achieved for the three ESPaDOnS observations. Marginal detection is achieved only for one Narval observation. No signal is detected in the null profiles (column 11). 

The global longitudinal fields for the Stokes $V$ LSD profiles (column 6) show a positive trend, hinting that the positive magnetic pole is located somewhere on the visible hemisphere.  Although the longitudinal field approaches zero,  the magnetic signal is still detectable given a sufficient s/n (for example 2007-03-06). This hints that we are looking nearly at the magnetic equator at some phases. 

\begin{figure}
	\includegraphics[width=84mm]{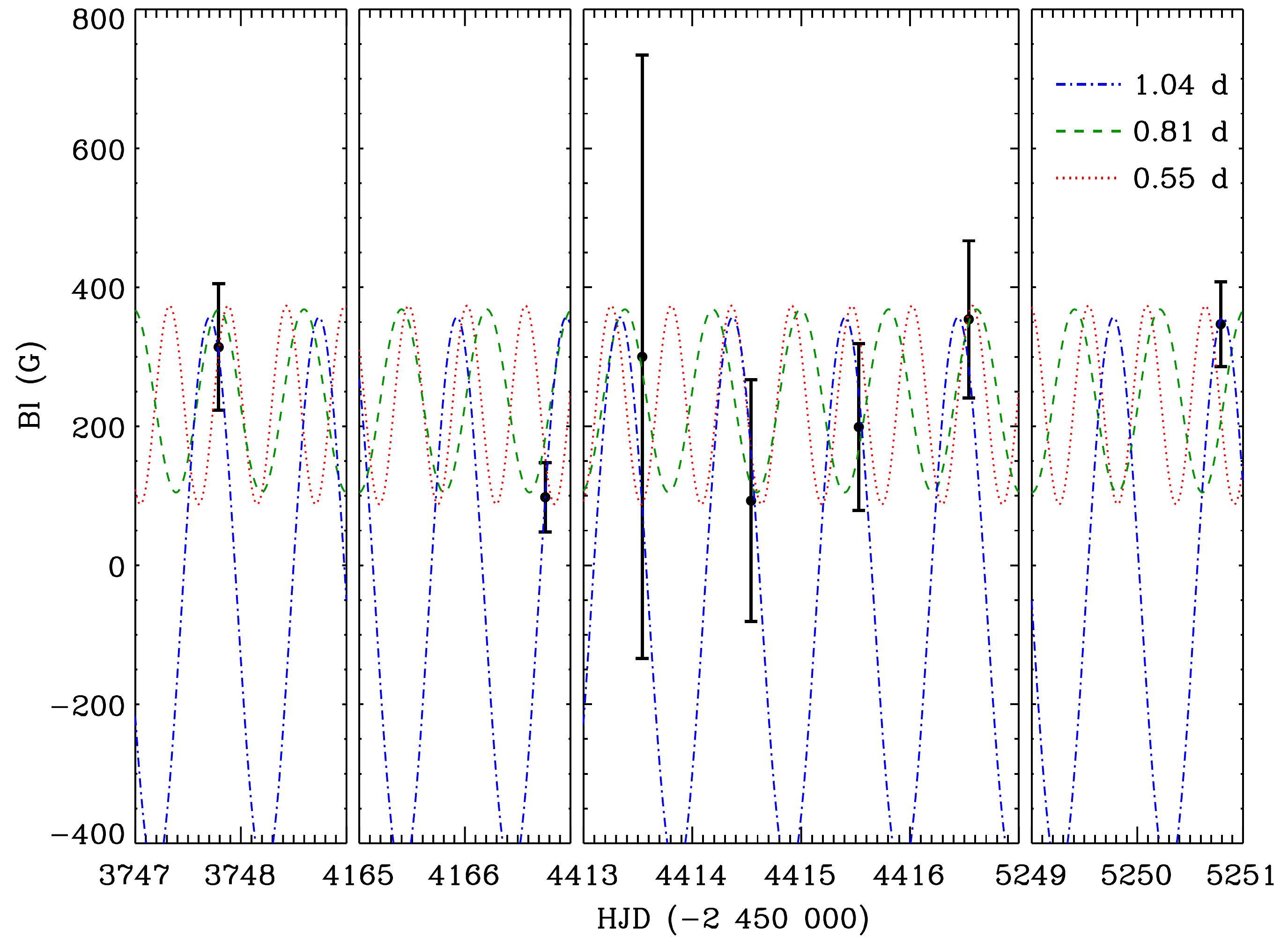}
	\caption{\label{fig|blsin} Longitudinal field curve for LP\,Ori (black dots) to which we have superposed sinusoidal curves for three of the possible periods. }
\end{figure}

Assuming a reasonable range of possible rotation periods, defined by the $v\sin i$ and breakup velocity, the longitudinal field curve and the LSD profiles can be phased with various periods. Figure \ref{fig|blsin} shows the longitudinal field measurements (black dots), along with sinusoidal curves for three possible periods. 
As no other variability has been observed by other means (spectral or photometric), this is a good example of a case where the rotational phases of the observations are not known. 

Given the range of global longitudinal field measurements, we extended the grid up to $B_p=3$\,kG. We set the $a$ parameter of the Jeffreys prior to 25\,G, corresponding to two times the parameter grid step. We performed the analysis on both the Stokes $V$ and null profiles. The fit of the Stokes $I$ profiles is shown in Figure \ref{fig|lpV}. Vertical dotted lines represent the velocity range of the fits. 

The odds ratios for each individual observation are given in Table \ref{tab|lp_odds}. The odds ratios for the null profiles (column 3) all favour the absence of a magnetic field ($M_0$ model). The odds ratios for the Stokes $V$ observations all favour the oblique dipole model ($M_1$), but vary from more than 10 orders of magnitude for the high s/n observations to less than one order of magnitude for the low s/n observations. As mentioned for realistic numerical tests (Section \ref{sec|real}), it is possible to obtain an odds ratio in favour of the dipole model when the noise happens to have a shape similar to a magnetic signal. When combining all the observations together, we get an odds ratio strongly in favour of the magnetic model, by 73 orders of magnitude, as expected given the definite signal detections in the ESPaDOnS observations. However, if only the low s/n observations were available (2007-11-08, -09 and -10), the situation would be more ambiguous. None of these observations lead to a traditional definite detection when considered on their own. We therefore combined and analysed these three observations in pairs, and then all together. The three possible pairs of observations led to odds ratios in favour of $M_1$ by more than one order of magnitude, and the combination of the three observations lead to an odds ratio $\log(M_0/M_1)=-3.3$, i.e. more than 3 orders of magnitudes. Therefore, the Bayesian algorithm is able to recover the magnetic signal that would have been buried in the noise and undetected by traditional diagnostics. 

\begin{table}
	\caption{\label{tab|lp_odds} Stokes $V$ and null profiles odds ratio for LP\,Ori observations taken individually, combined together, and for various combinations of the low s/n observations.}
	\begin{center}
	\begin{tabular}{l  c c}
	\hline
	\multicolumn{1}{c}{Date} & $\log(M_0/M_1)$  & $\log(M_0/M_1)$\\ 
		& $V$ & $N$ \\ 
		\multicolumn{1}{c}{(1)} & (2) & (3) \\ \hline
	Nar. 2007-11-08 	& -0.32 &0.15\\
	Nar. 2007-11-09  	&  -1.7 &0.18\\
	Nar. 2007-11-10  	& -0.86  &0.15\\
	Nar. 2007-11-11  	&  -11.5  &0.32\\
	ESP. 2006-01-11 	& -13.5 &  0.26 \\
	ESP. 2007-03-06 	& -24.8 & 0.42\\
	ESP. 2010-02-23 	 & -16.9 &0.43 \\ 
	Combination & -73.3 & 0.52\\
	2007-11 08+09 & -2.3 & 0.20 \\
	2007-11 08+10 & -1.3 & 0.17\\
	2007-11 09+10 & -2.7 & 0.16 \\
	2007-11 08+09+10 & -3.3 & 0.17 \\
	\hline
	\end{tabular}
	\end{center}
\end{table}

In Figures \ref{fig|lpV} and \ref{fig|lpN}, the grey dashed lines represent the $M_0$ model ($V=0$). The best fits achievable by a dipole model for each observation taken individually (maximum of the individual likelihoods) are shown in red. The best fit produced by a single $\mathcal{B}$ oblique dipole (maximum of the joint likelihood) is shown in green. The reduced $\chi^2$s are indicated with corresponding colours. 
For the Stokes $V$ observations where odds ratios strongly favour the $M_1$ model, the reduced $\chi^2$ is much improved with the addition of the more complex model compared to the $\chi^2$ of the $M_0$ model.
Not only are the data reproduced by the dipolar profiles, they can all be simultaneously reproduced by a single $\mathcal{B}$ configuration, as shown by the similar $\chi^2$ for the individual fit (red) and the maximum of the joint likelihood (green).  

However, the reduced $\chi^2$ remains high in one case (2010-02-23; $\chi^2_\mathrm{red}$=1.95), meaning that there is some extra variance in the observation that neither models are able to reproduce (a 3$\sigma$ deviation would correspond to a reduced $\chi^2$ of 1.65). The reduced $\chi^2$s are low for all the null profile observations (Figure \ref{fig|lpN}). This points toward a systematic deviation or a model-based effect for the Stokes $V$ observation of Feb. 2010 rather than an extra instrumental scatter or underestimated error bars.

\begin{figure}
\includegraphics[width=84mm]{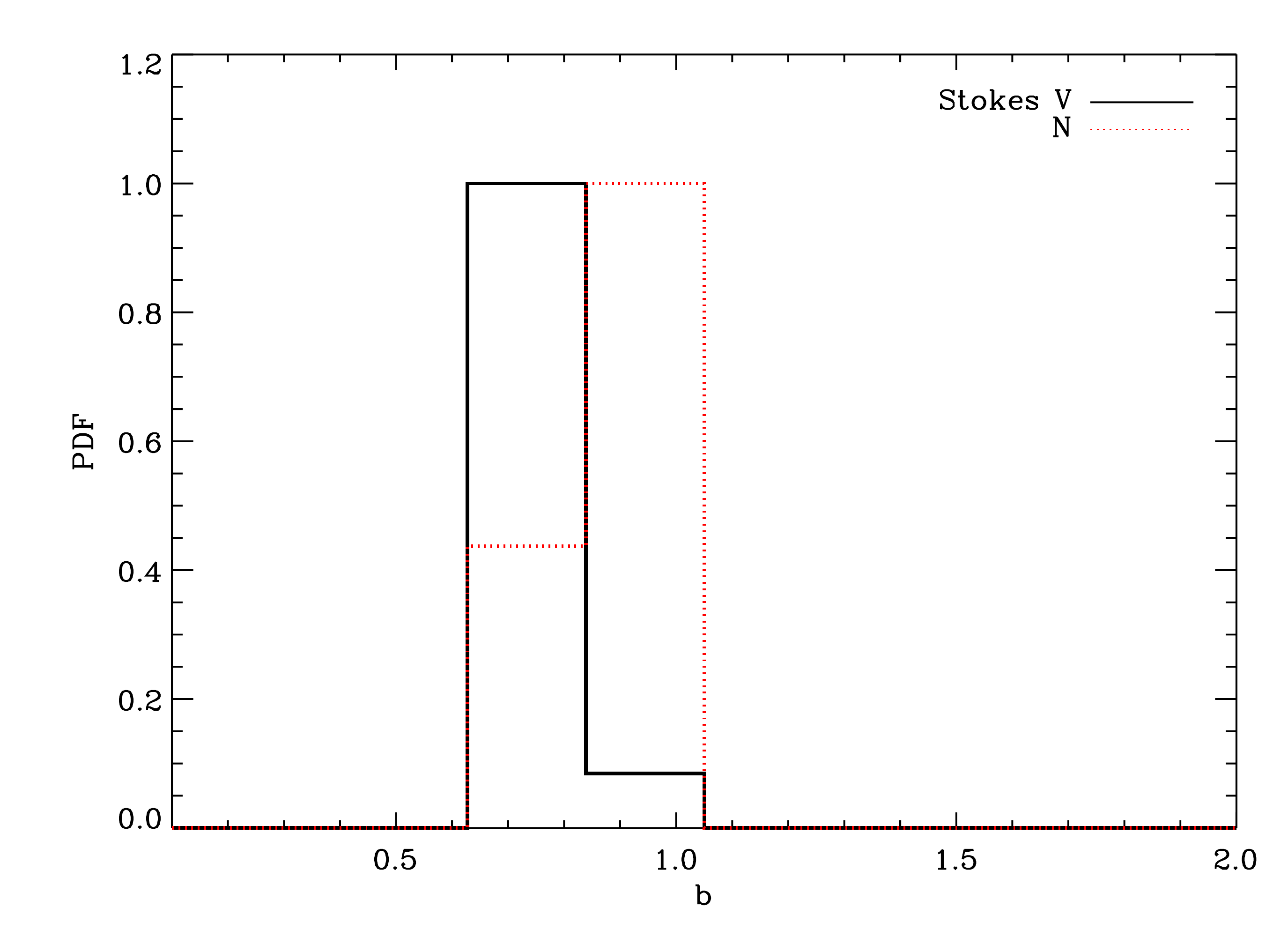}
\caption{\label{fig|scale_noise} Posterior probability density marginalised for the noise scaling parameter for the Stokes $V$ (black) and null $N$ profiles (dotted red) of LP\,Ori. }
\end{figure}

When performing parameter estimation, the extra scatter is addressed by the noise scaling term (Eq. \ref{eq|prior_b}). It is therefore possible to extract the probability density function marginalised for the noise scaling parameter, and determine if the model is able to reproduce the observations down to the noise level. Figure \ref{fig|scale_noise} shows that the Bayesian estimate of the variance is larger than the assumed variance (i.e. error bars) for both Stokes $V$ and the null profiles, but by less than a factor of two. 

\begin{figure}
\includegraphics[width=84mm]{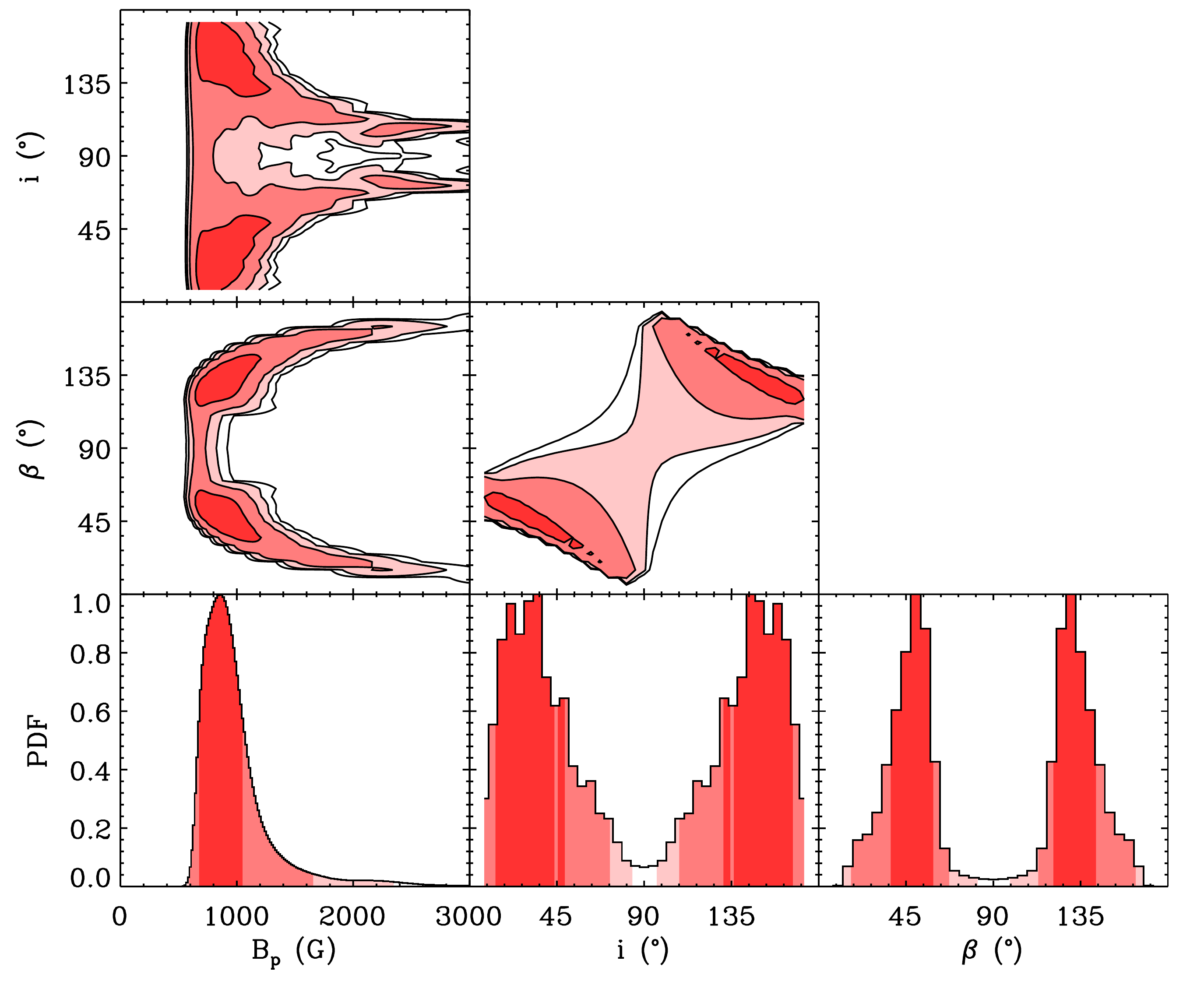}
\caption{\label{fig|lp_postV} Same as Figure \ref{fig|mar_tri_nodetect1} for the combined Stokes $V$ observations of LP\,Ori.}
\end{figure}

The posterior probability density for the Stokes $V$ observations is shown in Figure \ref{fig|lp_postV}. 
Dipole configurations with large inclination or obliquity are less favoured, as shown by the probability densities marginalised for $i$ and $\beta$, because the observations mainly show either the positive pole or the magnetic equator on the visible hemisphere. This is more likely to occur if the $i$ and $\beta$ angles are small and the negative pole spends little or no time on the visible hemisphere. 
The angle values are interrelated as shown by the $i$-$\beta$ 2D plane. If the inclination is small, the obliquity is more likely to be large, in order to display an equator-like magnetic signature. 
The probability density marginalised for the field strength shows a sharp lower limit. The high-$B_p$ tail of the distribution is attributable to the high-inclination (low obliquity) configurations, as shown in the 2D planes for $B_p$-$i$ and $B_p$-$\beta$. 

\begin{figure}
\includegraphics[width=84mm]{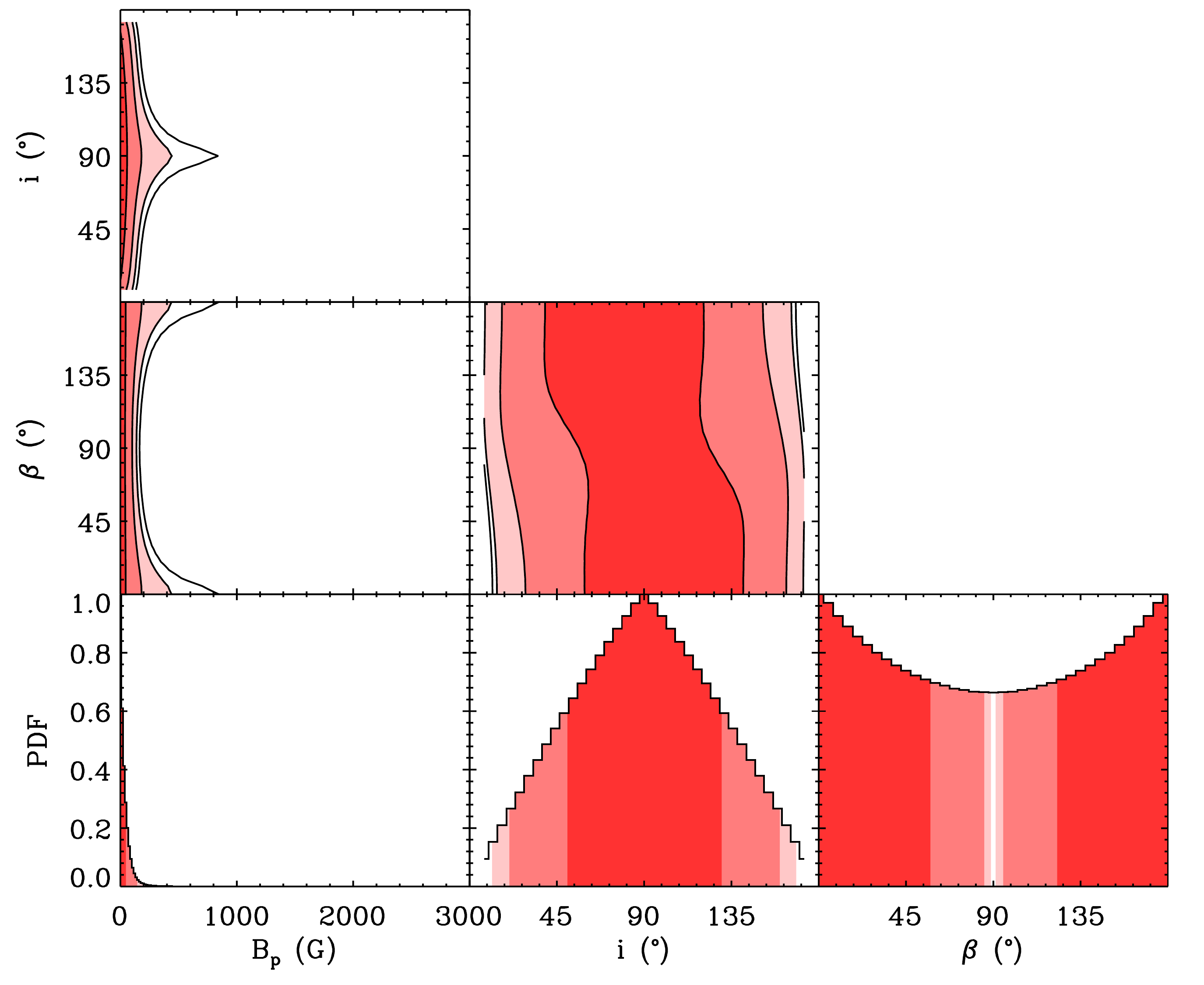}
\caption{\label{fig|lp_postN} Same as Figure \ref{fig|mar_tri_nodetect1} for the combined null $N$ observations of LP\,Ori.}
\end{figure}

\begin{figure}
\includegraphics[width=84mm]{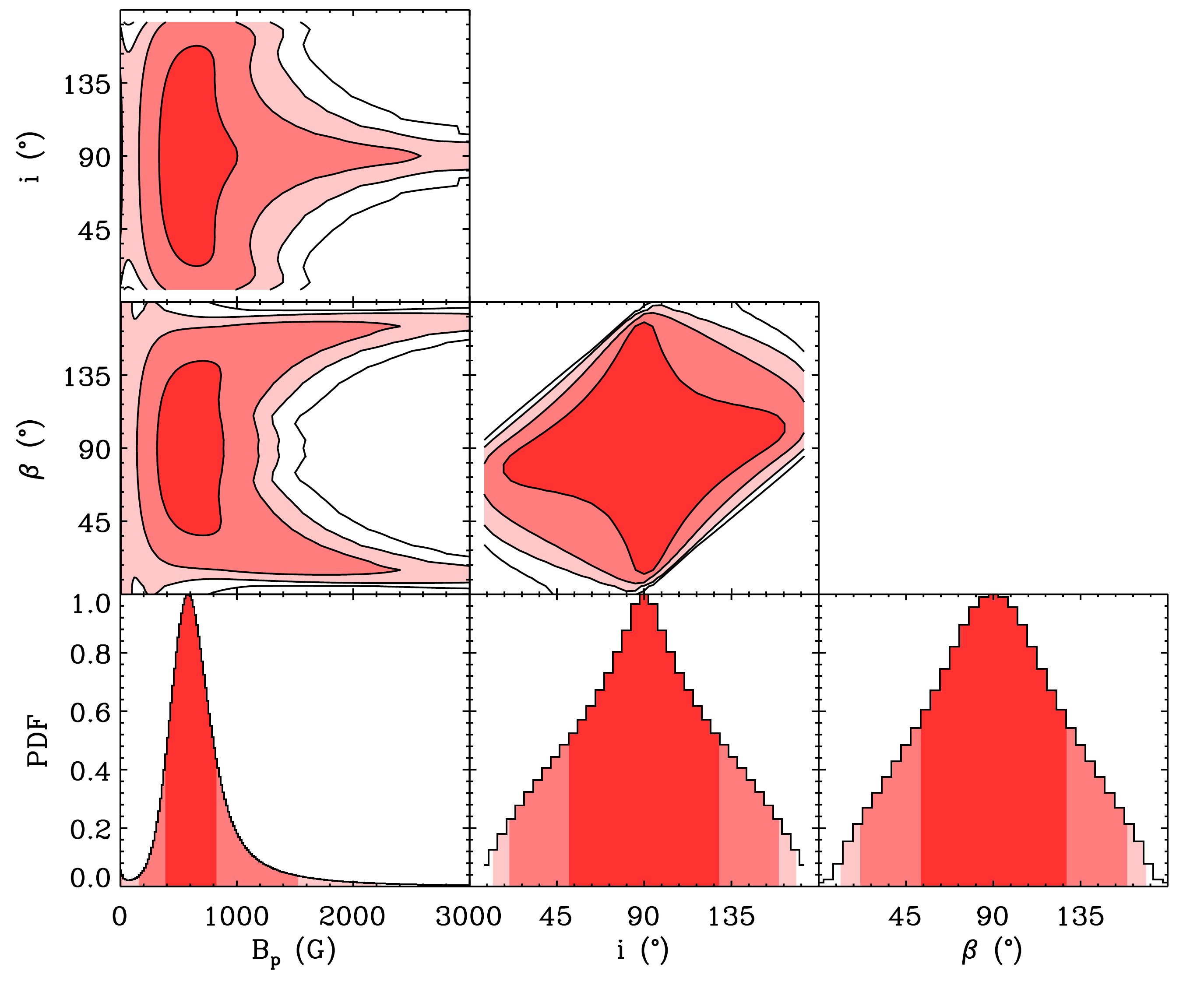}
\caption{\label{fig|lp_postV345} Same as Figure \ref{fig|mar_tri_nodetect1} for the combined low s/n Stokes $V$ observations of LP\,Ori (2007-11-08, 2007-08-09 and 2007-11-10). }
\end{figure}

Figure \ref{fig|lp_postN} shows the probability densities for the null profile observations, which are, as expected,  similar to the realistic simulation of pure noise (as shown in Figure \ref{fig|mar_tri_nodetect7}). 
We also show in Figure \ref{fig|lp_postV345} the probability densities we obtained when considering only the three low s/n observations. The constraints on the parameters, especially the angles, are worse than when we consider the full data set.  Therefore, the constraints on the angles are mainly defined by the high s/n observations. 

The credible regions of the probability density marginalised for $B_p$ are given in Table \ref{tab|lp_cred}. We also present the maximum of the joint posterior probability density (MAP), the mode of the probability density marginalised for $B_p$, as well as the median. 
For the Stokes $V$ observations, the MAP, mode and median values are all similar, therefore the fits to the data obtained for these values are nearly undistinguishable. The fit obtained with the MAP values is shown in Figure \ref{fig|lpV} (blue curve). Using the 68.3\% credible region and the median, the dipole field strength of LP\,Ori is estimated to be $911\stackrel{+138}{_{-244}}$\,G.  As was seen for the probability density in Figure \ref{fig|lp_postV}, the lower limit on the field strength is sharp. The 68.3\% credible region lower limit is 667\,G, and 557\,G for the 99.7\% credible region. The distribution has a high-$B_p$ tail, and the 99.7\% credible region extends up to 2.6\,kG. 
\begin{table*}
	\caption{\label{tab|lp_cred} Field strength estimation for Stokes $V$ and null profile observations of LP\,Ori. The first three columns give the maximum of the joint posterior probability density as well as the mode and the median of the joint posterior probability density marginalised for $B_p$. The last four columns give the credible regions. }
	\begin{tabular}{l c c c c c c c}
	\hline
	\multicolumn{1}{c}{Obs.} & MAP & Mode &  Median &  {68.3\%} & {95.4\%} & {99\%} & {99.7\%}\\ 
	& (G) & (G) & (G) & (G) & (G) & (G) & (G) \\
	\multicolumn{1}{c}{(1)} & (2) & (3) & (4) & (5) & {(6)} & {(7)} & {(8)}\\ \hline
	& \multicolumn{7}{c}{Stokes $V$}\\
	Combination 	&885&855&911& 667 - 1049 & 590 - 1657 & 572 - 2348 & 557 - 2653 \\
	No noise scaling &930&930&967& 762 - 1119 & 621 - 1659 & 660 - 2336 & 578 - 2672 \\
	Low s/n			&555&585&633& 386 - 825    & 0 - 1527    & 0 - 1141 & 0 - 2689 \\ \hline
	& \multicolumn{7}{c}{Null profiles}\\
	Combination	&0&0&27& 0 - 47 & 0 - 144 & 0 - 301 & 0 - 530 \\
	No noise scaling	&0&0&27& 0 - 46 & 0 - 142 & 0 - 296 & 0 - 524 \\
	Low s/n			&0&0&86& 0 - 163 & 0 - 502 & 0 - 1002 & 0 - 1624 \\
	\hline
	\end{tabular}
\end{table*}

The second row of Table \ref{tab|lp_cred} illustrates the effect of the noise scaling parameter on the inferred field values. When considering the variance as a model parameter, the MAP, mode and median are shifted to slightly lower $B_p$ values. The credible regions are somewhat larger, and also slightly shifted to lower values. 

The parameter estimation for the low s/n observations is less robust than that of the full dataset. The estimated field value from the median and the 68.3\% credible region is $633\stackrel{+192}{_{-247}}$\,G. However, we do not have a good constraint on the lower limit, as the credible regions quickly go to zero. The 99.7\% credible region extends to 2.7\,kG as well.

For the null profile observations, the MAPs and modes are all 0\,G. Given the shape of the probability distribution, the median is non-null. However, given the probability density in favour of the $M_0$ model, it makes more sense to express the field strength estimation in terms of the upper limit of the credible regions. The null profiles are a good representation of what our data would look like in the absence of a stellar magnetic field. For the whole $N$ dataset, the upper limit of the 95.4\% credible region is 144\,G, well below the inferred dipole strength ($\sim1$\,kG) from the Stokes $V$ observations. For the 99.0\% credible region, we can say that the probability that an undetected field would have a dipole strength of more than 300\,G is only 1\%. 
The noise scaling does not significantly change the credible regions. 
Interestingly, the dipole strength inferred from the low s/n Stokes $V$ observations is around 600\,G, and the odds ratio favours the magnetic model. From the low s/n null $N$ profile observation, we can see that a field with a dipole strength of the order of the 95.4\% credible region upper limit (502\,G) can indeed be detected.

\section{Conclusion}
\label{sec|conclu}

In this paper, we have described a method based on Bayesian statistics to infer the magnetic properties of stars observed spectropolarimetrically in the context of large surveys like the Magnetism in Massive Stars project. This approach is well-suited for stars for which the stellar rotation period, and therefore the rotational phases of the small number of observations, are not known.

The model used to predict the expected Stokes $V$ profiles is that of an oblique dipolar magnetic field, parametrised by the field strength at the pole, the inclination of the rotational axis, the obliquity of the magnetic axis with respect to the rotational axis and the rotational phase (which is allowed to take any value for each individual observation). In the present case, the calculations are performed under the weak-field approximation, although any polarised spectral synthesis code can in principle be used with the Bayesian algorithm.

The result of the analysis is a multidimensional posterior probability density that describes the relative likelihood of models spanning the parameter space of the dipolar field model.  We have used synthetic observations to explore the behaviour of the Bayesian algorithm under ideal and realistic conditions. In the case of an ideal non-detection, the posterior probability density for the field strength has the form of a decreasing exponential, the extended tail of the field strength distribution being due to a specific family of dipole orientations. This tail becomes less extended with the addition of multiple observations, as some of these specific dipole orientations only present low-amplitude or null Stokes $V$ profiles over a restricted range of phases. However, the possibility that the dipole is aligned with the rotational axis and seen equator-on ($i=90^\circ$ and $\beta=0^\circ$) always remains as this configuration never produces any circular polarisation at any phase. 
When a detectable field is present, the probability distribution shows a sharp lower limit on the dipolar field strength, and a slow decrease towards higher field strengths, producing an asymmetrical distribution.  With only one observation, not much can be inferred about the dipole orientation. The longitudinal field indicates which pole is located on the visible hemisphere and the shape of the Stokes V profile indicates when some obliquity is necessary because the pole is located at a non-null rotational radial velocity.

A particularly useful quantity that can be computed from the posterior probabilities is the so-called ``odds ratio'', which compares the relative compatibility between the observations and the magnetic dipole hypothesis versus the non-magnetic hypothesis. 
By adding a magnetic signal corresponding to the upper limit of the credible regions found for the ideal non-detection, we have explored the detection capability of the odds ratios. We find that fields corresponding to the 68.3\% and 95.4\% region are below detection, whereas those corresponding to the 99.0\% and 99.7\% regions are well detected with the odds ratios. In contrast, traditional diagnostics (detection probability and global longitudinal field significance) only detect fields corresponding to the 99.7\% region.

By using a set of five realistic simulated observations, we have also shown that in the case of noise emulating the shape of a weak magnetic signal, it is possible to use the odds ratios to distinguish between noise and real signal by obtaining a small number of additional observations. Combining all the observations together, it is therefore possible to detect a weak magnetic field (with a strength corresponding to roughly the upper limit of the 95.4\% credible region of the noise-only case) under the detection capabilities of the traditional diagnostics. 
We have therefore shown that the odds ratio is an powerful quantitative indicator of which undetected stars in a survey should be re-observed in priority.

In most applications where a field is indeed detected, the resulting probability density will generally be marginalised for the dipole strength, as only limited information can be obtained for the rotational inclination and the magnetic obliquity from a few observations (the magnetic geometry is recovered by the most probable inclination and obliquity, but the credible regions are quite extended). We have shown that the dipole strength probability distribution provides a reasonable estimate of the field strength.

We have applied our method to real spectropolarimetric observations of the magnetic B-type star LP\,Ori. The dataset consists of 3 ESPaDOnS and 4 Narval observations, of various signal-to-noise ratios.   
A magnetic signal is indeed detected by the odds ratios, even when only considering the low s/n observations where the traditional diagnostics do not detect the magnetic field. Using all the available spectra, we used the median of the marginalised posterior probability density, as well as the 68.3\% credible region, to infer a dipolar field strength of $911\stackrel{+138}{_{-244}}$\,G. 
Although the probability density for the obliquity and inclination do not provide any tight constraints, geometries for which the negative magnetic pole spends little or no time on the visible hemisphere are preferred, since the dataset consists of positive or nearly null longitudinal field measurements.
We also performed our analysis on the diagnostic null spectra, which resulted in a non-detection. The null profiles provide an useful verification of spurious signals and also provide an estimate of the field strengths that would be detectable with data of such quality. For example, when considering only the low s/n observations, the Stokes $V$ profile analysis yielded the detection of the $\sim600$\,G dipolar field and the diagnostic null profile analysis yielded an upper limit of $\sim500$\,G for the 95.4\% credible region.

\section*{Acknowledgments}
  VP acknowledges support from Fonds qu\'eb\'ecois de la recherche sur la nature et les technologies.
    GAW  acknowledges support from the Discovery Grants programme of the Natural Science and Engineering Research Council of Canada.
 We thank M. Gagn\'e and R.~P. Breton for useful discussions, and the anonymous referee whose helpful comments led to the improvement of this paper. 
\bibliographystyle{mn2e}
\bibliography{bayes}
\clearpage

\begin{table*}
\small
\caption{\label{tab|mask}  Line mask used for the LSD profile construction of LP\,Ori. The given values are the central wavelength, the ion, the unbroaden line depth with respect to the continuum and the effective Land\'e factor.}
\begin{tabular}{cccc cccc cccc }
\hline\hline
$\lambda_0$ & Ion & $d$ & $g_\mathrm{eff}$ & $\lambda_0$ & Ion & $d$ & $g_\mathrm{eff}$ & $\lambda_0$ & Ion & $d$ & $g_\mathrm{eff}$  \\
(nm) & & ($I_c$) & &(nm) & & ($I_c$) & &(nm) & & ($I_c$) & \\
(1) & (2) & (3) &(4) &(1) & (2) & (3) &(4) &(1) & (2) & (3) &(4) \\\hline
 360.0943 & Fe\,\textsc{iii} & 0.159 & 1.075  & 	  437.1337 & Fe\,\textsc{iii} & 0.062 & 1.480  & 	  480.6021 & Ar\,\textsc{ii} & 0.048 & 1.600 \\
  360.1630 & Al\,\textsc{iii} & 0.214 & 1.100  & 	  437.2331 & C\,\textsc{ii} & 0.042 & 0.800  & 	  481.3333 & Si\,\textsc{iii} & 0.032 & 0.833 \\
  360.3890 & Fe\,\textsc{iii} & 0.156 & 0.670  & 	  437.2375 & C\,\textsc{ii} & 0.042 & 1.467  & 	  481.5552 & S\,\textsc{ii} & 0.100 & 1.300 \\
  386.7470 & He\,\textsc{i} & 0.408 & 1.250  & 	  437.2823 & Fe\,\textsc{iii} & 0.043 & 1.155  & 	  482.8951 & Si\,\textsc{iii} & 0.091 & 1.100 \\
  386.7482 & He\,\textsc{i} & 0.356 & 1.750  & 	  437.4281 & C\,\textsc{ii} & 0.058 & 1.214  & 	  494.2473 & S\,\textsc{ii} & 0.112 & 2.667 \\
  386.7630 & He\,\textsc{i} & 0.225 & 2.000  & 	  437.6582 & C\,\textsc{ii} & 0.042 & 1.100  & 	  500.1135 & N\,\textsc{ii} & 0.053 & 0.750 \\
  387.1791 & He\,\textsc{i} & 0.430 & 1.000  & 	  439.5755 & Fe\,\textsc{iii} & 0.181 & 1.515  & 	  500.1475 & N\,\textsc{ii} & 0.062 & 1.000 \\
  387.6038 & C\,\textsc{ii} & 0.054 & 0.700  & 	  440.9990 & C\,\textsc{ii} & 0.050 & 1.286  & 	  500.1959 & Fe\,\textsc{ii} & 0.073 & 1.150 \\
  387.6187 & C\,\textsc{ii} & 0.076 & 1.136  & 	  441.1152 & C\,\textsc{ii} & 0.052 & 0.900  & 	  501.4042 & S\,\textsc{ii} & 0.073 & 1.333 \\
  387.6393 & C\,\textsc{ii} & 0.069 & 1.056  & 	  441.1510 & C\,\textsc{ii} & 0.063 & 1.071  & 	  501.5678 & He\,\textsc{i} & 0.838 & 1.000 \\
  387.6653 & C\,\textsc{ii} & 0.061 & 0.929  & 	  441.4900 & O\,\textsc{ii} & 0.171 & 1.100  & 	  501.8440 & Fe\,\textsc{ii} & 0.035 & 1.935 \\
  387.8181 & He\,\textsc{i} & 0.045 & 1.000  & 	  441.6979 & O\,\textsc{ii} & 0.145 & 0.833  & 	  503.2126 & C\,\textsc{ii} & 0.057 & 1.100 \\
  391.1962 & O\,\textsc{ii} & 0.234 & 1.100  & 	  441.9596 & Fe\,\textsc{iii} & 0.166 & 1.665  & 	  503.2434 & S\,\textsc{ii} & 0.122 & 1.600 \\
  394.5034 & O\,\textsc{ii} & 0.336 & 1.500  & 	  443.7551 & He\,\textsc{i} & 0.471 & 1.000  & 	  503.5708 & Fe\,\textsc{ii} & 0.047 & 1.222 \\
  399.4997 & N\,\textsc{ii} & 0.346 & 1.000  & 	  444.7030 & N\,\textsc{ii} & 0.136 & 1.000  & 	  503.5946 & C\,\textsc{ii} & 0.040 & 0.833 \\
  399.7926 & Si\,\textsc{ii} & 0.106 & 0.929  & 	  451.2565 & Al\,\textsc{iii} & 0.132 & 0.833  & 	  504.1024 & Si\,\textsc{ii} & 0.130 & 0.833 \\
  403.9160 & Fe\,\textsc{iii} & 0.102 & 0.670  & 	  452.4675 & S\,\textsc{ii} & 0.044 & 1.067  & 	  504.5103 & N\,\textsc{ii} & 0.054 & 1.250 \\
  406.9621 & O\,\textsc{ii} & 0.097 & 0.500  & 	  452.4941 & S\,\textsc{ii} & 0.044 & 1.100  & 	  504.7738 & He\,\textsc{i} & 0.503 & 1.000 \\
  406.9881 & O\,\textsc{ii} & 0.113 & 0.900  & 	  452.8945 & Al\,\textsc{iii} & 0.060 & 1.067  & 	  507.3903 & Fe\,\textsc{iii} & 0.099 & 2.010 \\
  407.2005 & Ar\,\textsc{ii} & 0.063 & 1.200  & 	  452.9189 & Al\,\textsc{iii} & 0.155 & 1.100  & 	  512.2272 & C\,\textsc{ii} & 0.122 & 1.071 \\
  407.2150 & O\,\textsc{ii} & 0.126 & 1.071  & 	  454.9474 & Fe\,\textsc{ii} & 0.028 & 1.035  & 	  512.7387 & Fe\,\textsc{iii} & 0.068 & 1.180 \\
  407.4480 & C\,\textsc{ii} & 0.059 & 0.500  & 	  455.2410 & S\,\textsc{ii} & 0.125 & 0.833  & 	  512.7631 & Fe\,\textsc{iii} & 0.047 & 1.675 \\
  407.4543 & C\,\textsc{ii} & 0.078 & 0.900  & 	  455.2622 & Si\,\textsc{iii} & 0.259 & 1.250  & 	  513.2950 & C\,\textsc{ii} & 0.092 & 1.500 \\
  407.4841 & C\,\textsc{ii} & 0.096 & 1.071  & 	  456.7840 & Si\,\textsc{iii} & 0.313 & 1.750  & 	  513.3280 & C\,\textsc{ii} & 0.095 & 1.500 \\
  407.5852 & C\,\textsc{ii} & 0.114 & 1.167  & 	  457.4757 & Si\,\textsc{iii} & 0.232 & 2.000  & 	  514.3497 & C\,\textsc{ii} & 0.075 & 1.500 \\
  407.5859 & O\,\textsc{ii} & 0.138 & 1.167  & 	  459.0974 & O\,\textsc{ii} & 0.149 & 1.071  & 	  514.5167 & C\,\textsc{ii} & 0.144 & 1.600 \\
  407.8839 & O\,\textsc{ii} & 0.066 & 0.800  & 	  459.6172 & O\,\textsc{ii} & 0.131 & 0.900  & 	  515.1085 & C\,\textsc{ii} & 0.121 & 1.500 \\
  408.1007 & Fe\,\textsc{iii} & 0.071 & 1.375  & 	  460.1481 & N\,\textsc{ii} & 0.180 & 1.500  & 	  515.6111 & Fe\,\textsc{iii} & 0.117 & 1.245 \\
  412.8054 & Si\,\textsc{ii} & 0.118 & 0.900  & 	  460.7149 & N\,\textsc{ii} & 0.124 & 1.500  & 	  516.9033 & Fe\,\textsc{ii} & 0.060 & 1.325 \\
  413.0872 & Si\,\textsc{ii} & 0.048 & 1.029  & 	  460.9567 & Ar\,\textsc{ii} & 0.083 & 1.071  & 	  524.3306 & Fe\,\textsc{iii} & 0.085 & 1.300 \\
  413.0894 & Si\,\textsc{ii} & 0.110 & 1.071  & 	  461.3868 & N\,\textsc{ii} & 0.076 & 1.500  & 	  524.7952 & Fe\,\textsc{ii} & 0.030 & 0.735 \\
  413.2804 & O\,\textsc{ii} & 0.127 & 1.500  & 	  461.8559 & C\,\textsc{ii} & 0.103 & 0.929  & 	  545.3855 & S\,\textsc{ii} & 0.193 & 1.214 \\
  416.2665 & S\,\textsc{ii} & 0.194 & 1.167  & 	  461.9249 & C\,\textsc{ii} & 0.114 & 1.056  & 	  547.3614 & S\,\textsc{ii} & 0.094 & 1.333 \\
  416.4731 & Fe\,\textsc{iii} & 0.017 & 1.300  & 	  462.1396 & N\,\textsc{ii} & 0.092 & 1.500  & 	  563.9977 & S\,\textsc{ii} & 0.179 & 1.100 \\
  416.6840 & Fe\,\textsc{iii} & 0.010 & 1.700  & 	  462.5639 & C\,\textsc{ii} & 0.053 & 1.143  & 	  564.0346 & S\,\textsc{ii} & 0.048 & 1.071 \\
  417.4265 & S\,\textsc{ii} & 0.138 & 1.056  & 	  463.0543 & N\,\textsc{ii} & 0.243 & 1.500  & 	  564.5681 & S\,\textsc{ii} & 0.011 & 1.900 \\
  418.5440 & O\,\textsc{ii} & 0.152 & 0.929  & 	  463.8851 & O\,\textsc{ii} & 0.118 & 0.833  & 	  564.7020 & S\,\textsc{ii} & 0.153 & 0.833 \\
  418.9681 & S\,\textsc{ii} & 0.067 & 1.200  & 	  464.1810 & O\,\textsc{ii} & 0.160 & 1.100  & 	  569.6604 & Al\,\textsc{iii} & 0.158 & 1.167 \\
  418.9789 & O\,\textsc{ii} & 0.062 & 1.056  & 	  464.3090 & N\,\textsc{ii} & 0.129 & 1.500  & 	  572.2730 & Al\,\textsc{iii} & 0.139 & 1.333 \\
  419.0707 & Si\,\textsc{ii} & 0.064 & 1.100  & 	  464.9138 & O\,\textsc{ii} & 0.243 & 1.214  & 	  573.9734 & Si\,\textsc{iii} & 0.221 & 1.000 \\
  425.3589 & S\,\textsc{iii} & 0.192 & 1.167  & 	  465.0838 & O\,\textsc{ii} & 0.152 & 1.333  & 	  583.3938 & Fe\,\textsc{iii} & 0.073 & 1.375 \\
  426.6999 & C\,\textsc{ii} & 0.268 & 0.900  & 	  465.6757 & S\,\textsc{ii} & 0.034 & 1.833  & 	  587.5599 & He\,\textsc{i} & 0.177 & 2.000 \\
  426.7259 & C\,\textsc{ii} & 0.173 & 1.029  & 	  466.1635 & O\,\textsc{ii} & 0.135 & 1.467  & 	  587.5614 & He\,\textsc{i} & 0.248 & 1.333 \\
  426.7259 & C\,\textsc{ii} & 0.277 & 1.071  & 	  466.3046 & Al\,\textsc{ii} & 0.086 & 1.000  & 	  587.5615 & He\,\textsc{i} & 0.281 & 1.167 \\
  426.7762 & S\,\textsc{ii} & 0.111 & 1.100  & 	  467.6231 & O\,\textsc{ii} & 0.185 & 1.486  & 	  587.5625 & He\,\textsc{i} & 0.248 & 1.000 \\
  430.4767 & Fe\,\textsc{iii} & 0.142 & 1.150  & 	  469.9215 & O\,\textsc{ii} & 0.135 & 0.900  & 	  587.5640 & He\,\textsc{i} & 0.270 & 1.000 \\
  431.0355 & Fe\,\textsc{iii} & 0.084 & 1.220  & 	  470.5343 & O\,\textsc{ii} & 0.129 & 1.071  & 	  587.5966 & He\,\textsc{i} & 0.254 & 0.500 \\
  431.7136 & O\,\textsc{ii} & 0.081 & 1.500  & 	  471.3139 & He\,\textsc{i} & 0.310 & 1.250  & 	  637.1371 & Si\,\textsc{ii} & 0.176 & 1.333 \\
  431.7266 & C\,\textsc{ii} & 0.052 & 1.600  & 	  471.3156 & He\,\textsc{i} & 0.291 & 1.750  & 	  667.8154 & He\,\textsc{i} & 1.113 & 1.000 \\
  431.8643 & S\,\textsc{ii} & 0.061 & 1.486  & 	  471.3376 & He\,\textsc{i} & 0.238 & 2.000  & 	  728.1349 & He\,\textsc{i} & 0.524 & 1.000 \\
  431.9625 & O\,\textsc{ii} & 0.082 & 1.500  & 	  471.6271 & S\,\textsc{ii} & 0.025 & 1.867  & 	&& \\
  436.6893 & O\,\textsc{ii} & 0.073 & 1.500 &	  480.3288 & N\,\textsc{ii} & 0.113 & 1.333  & 	 &&\\
 
\hline
\end{tabular}
\end{table*}

\label{lastpage}

\end{document}

%% file: aas_macros.tex
%
%
%


\def\jnl@style{\it}
\def\aaref@jnl#1{{\jnl@style#1}}

\def\aaref@jnl#1{{\jnl@style#1}}

\def\aj{\aaref@jnl{AJ}}                   
\def\araa{\aaref@jnl{ARA\&A}}             
\def\apj{\aaref@jnl{ApJ}}                 
\def\apjl{\aaref@jnl{ApJ}}                
\def\apjs{\aaref@jnl{ApJS}}               
\def\ao{\aaref@jnl{Appl.~Opt.}}           
\def\apss{\aaref@jnl{Ap\&SS}}             
\def\aap{\aaref@jnl{A\&A}}                
\def\aapr{\aaref@jnl{A\&A~Rev.}}          
\def\aaps{\aaref@jnl{A\&AS}}              
\def\azh{\aaref@jnl{AZh}}                 
\def\baas{\aaref@jnl{BAAS}}               
\def\jrasc{\aaref@jnl{JRASC}}             
\def\memras{\aaref@jnl{MmRAS}}            
\def\mnras{\aaref@jnl{MNRAS}}             
\def\pra{\aaref@jnl{Phys.~Rev.~A}}        
\def\prb{\aaref@jnl{Phys.~Rev.~B}}        
\def\prc{\aaref@jnl{Phys.~Rev.~C}}        
\def\prd{\aaref@jnl{Phys.~Rev.~D}}        
\def\pre{\aaref@jnl{Phys.~Rev.~E}}        
\def\prl{\aaref@jnl{Phys.~Rev.~Lett.}}    
\def\pasp{\aaref@jnl{PASP}}               
\def\pasj{\aaref@jnl{PASJ}}               
\def\qjras{\aaref@jnl{QJRAS}}             
\def\skytel{\aaref@jnl{S\&T}}             
\def\solphys{\aaref@jnl{Sol.~Phys.}}      
\def\sovast{\aaref@jnl{Soviet~Ast.}}      
\def\ssr{\aaref@jnl{Space~Sci.~Rev.}}     
\def\zap{\aaref@jnl{ZAp}}                 
\def\nat{\aaref@jnl{Nature}}              
\def\iaucirc{\aaref@jnl{IAU~Circ.}}       
\def\aplett{\aaref@jnl{Astrophys.~Lett.}} 
\def\apspr{\aaref@jnl{Astrophys.~Space~Phys.~Res.}}
\def\bain{\aaref@jnl{Bull.~Astron.~Inst.~Netherlands}} 
\def\fcp{\aaref@jnl{Fund.~Cosmic~Phys.}}  
\def\gca{\aaref@jnl{Geochim.~Cosmochim.~Acta}}   
\def\grl{\aaref@jnl{Geophys.~Res.~Lett.}} 
\def\jcp{\aaref@jnl{J.~Chem.~Phys.}}      
\def\jgr{\aaref@jnl{J.~Geophys.~Res.}}    
\def\jqsrt{\aaref@jnl{J.~Quant.~Spec.~Radiat.~Transf.}}
\def\memsai{\aaref@jnl{Mem.~Soc.~Astron.~Italiana}}
\def\nphysa{\aaref@jnl{Nucl.~Phys.~A}}   
\def\physrep{\aaref@jnl{Phys.~Rep.}}   
\def\physscr{\aaref@jnl{Phys.~Scr}}   
\def\planss{\aaref@jnl{Planet.~Space~Sci.}}   
\def\procspie{\aaref@jnl{Proc.~SPIE}}   

\let\astap=\aap
\let\apjlett=\apjl
\let\apjsupp=\apjs
\let\applopt=\ao

%% file: bayes.bbl
\begin{thebibliography}{}

\bibitem[\protect\citeauthoryear{{Abt}, {Wang} \& {Cardona}}{{Abt}
  et~al.}{1991}]{1991ApJ...367..155A}
{Abt} H.~A.,  {Wang} R.,    {Cardona} O.,  1991, \apj, 367, 155

\bibitem[\protect\citeauthoryear{{Arregui} \& {Asensio Ramos}}{{Arregui} \&
  {Asensio Ramos}}{2011}]{2011ApJ...740...44A}
{Arregui} I.,  {Asensio Ramos} A.,  2011, \apj, 740, 44

\bibitem[\protect\citeauthoryear{{Asensio Ramos}}{{Asensio
  Ramos}}{2011}]{2011ApJ...731...27A}
{Asensio Ramos} A.,  2011, \apj, 731, 27

\bibitem[\protect\citeauthoryear{{Auri{\`e}re}, {Wade}, {Silvester},
  {Ligni{\`e}res}, {Bagnulo}, {Bale}, {Dintrans}, {Donati}, {Folsom} \&
  {Gruberbauer}}{{Auri{\`e}re} et~al.}{2007}]{2007A&A...475.1053A}
{Auri{\`e}re} M.,  {Wade} G.~A.,  {Silvester} J.,  {Ligni{\`e}res} F.,
  {Bagnulo} S.,  {Bale} K.,  {Dintrans} B.,  {Donati} J.~F., et~al.,  2007, \aap, 475, 1053

\bibitem[\protect\citeauthoryear{{Babel} \& {Montmerle}}{{Babel} \&
  {Montmerle}}{1997a}]{1997ApJ...485L..29B}
{Babel} J.,  {Montmerle} T.,  1997a, \apjl, 485, L29

\bibitem[\protect\citeauthoryear{{Babel} \& {Montmerle}}{{Babel} \&
  {Montmerle}}{1997b}]{1997A&A...323..121B}
{Babel} J.,  {Montmerle} T.,  1997b, \aap, 323, 121

\bibitem[\protect\citeauthoryear{{Bagnulo}, {Landstreet}, {Mason}, {Andretta},
  {Silaj} \& {Wade}}{{Bagnulo} et~al.}{2006}]{2006A&A...450..777B}
{Bagnulo} S.,  {Landstreet} J.~D.,  {Mason} E.,  {Andretta} V.,  {Silaj} J.,
  {Wade} G.~A.,  2006, \aap, 450, 777

\bibitem[\protect\citeauthoryear{{Bagnulo}, {Szeifert}, {Wade}, {Landstreet} \&
  {Mathys}}{{Bagnulo} et~al.}{2002}]{2002A&A...389..191B}
{Bagnulo} S.,  {Szeifert} T.,  {Wade} G.~A.,  {Landstreet} J.~D.,    {Mathys}
  G.,  2002, \aap, 389, 191

\bibitem[\protect\citeauthoryear{{Borra}, {Fletcher} \& {Poeckert}}{{Borra}
  et~al.}{1981}]{1981ApJ...247..569B}
{Borra} E.~F.,  {Fletcher} J.~M.,    {Poeckert} R.,  1981, \apj, 247, 569

\bibitem[\protect\citeauthoryear{{Borra} \& {Landstreet}}{{Borra} \&
  {Landstreet}}{1980}]{1980ApJS...42..421B}
{Borra} E.~F.,  {Landstreet} J.~D.,  1980, \apjs, 42, 421

\bibitem[\protect\citeauthoryear{{Donati}, {Semel}, {Carter}, {Rees} \&
  {Collier Cameron}}{{Donati} et~al.}{1997}]{1997MNRAS.291..658D}
{Donati} J.-F.,  {Semel} M.,  {Carter} B.~D.,  {Rees} D.~E.,    {Collier
  Cameron} A.,  1997, \mnras, 291, 658

\bibitem[\protect\citeauthoryear{{Gray}}{{Gray}}{1992}]{1992oasp.book.....G}
{Gray} D.~F.,  1992, {The Observation and Analysis of Stellar Photospheres.}.
Camb.~Astrophys.~Ser., Vol.~20

\bibitem[\protect\citeauthoryear{{Gregory}}{{Gregory}}{2005a}]{2005ApJ...631.1198G}
{Gregory} P.~C.,  2005a, \apj, 631, 1198

\bibitem[\protect\citeauthoryear{{Gregory}}{{Gregory}}{2005b}]{2005blda.book.....G}
{Gregory} P.~C.,  2005b, {Bayesian Logical Data Analysis for the Physical
  Sciences: A Comparative Approach with `Mathematica' Support}.
Cambridge University Press, Cambridge, UK.

\bibitem[\protect\citeauthoryear{{Hunter}, {Lennon}, {Dufton}, {Trundle},
  {Sim{\'o}n-D{\'{\i}}az}, {Smartt}, {Ryans} \& {Evans}}{{Hunter}
  et~al.}{2008}]{2008A&A...479..541H}
{Hunter} I.,  {Lennon} D.~J.,  {Dufton} P.~L.,  {Trundle} C.,
  {Sim{\'o}n-D{\'{\i}}az} S.,  {Smartt} S.~J.,  {Ryans} R.~S.~I.,    {Evans}
  C.~J.,  2008, \aap, 479, 541

\bibitem[\protect\citeauthoryear{{Jaynes}}{{Jaynes}}{2003}]{Jaynes2003}
{Jaynes} E.~T.,  2003, {Probability Theory: The Logic of Science}.
{Cambridge University Press, Cambridge, UK}

\bibitem[\protect\citeauthoryear{{Jeffreys}}{{Jeffreys}}{1998}]{Jeffreys1998}
{Jeffreys} H.,  1998, Theory of Probability; 3 edition.
Oxford University Press, Oxford, UK

\bibitem[\protect\citeauthoryear{{Kochukhov}, {Makaganiuk} \&
  {Piskunov}}{{Kochukhov} et~al.}{2010}]{2010A&A...524A...5K}
{Kochukhov} O.,  {Makaganiuk} V.,    {Piskunov} N.,  2010, \aap, 524, A5

\bibitem[\protect\citeauthoryear{{Kochukhov} \& {Wade}}{{Kochukhov} \&
  {Wade}}{2010}]{2010A&A...513A..13K}
{Kochukhov} O.,  {Wade} G.~A.,  2010, \aap, 513, A13

\bibitem[\protect\citeauthoryear{{Kupka}, {Ryabchikova}, {Piskunov}, {Stempels}
  \& {Weiss}}{{Kupka} et~al.}{2000}]{2000BaltA...9..590K}
{Kupka} F.~G.,  {Ryabchikova} T.~A.,  {Piskunov} N.~E.,  {Stempels} H.~C.,
  {Weiss} W.~W.,  2000, Baltic Astronomy, 9, 590

\bibitem[\protect\citeauthoryear{{Landi degl'Innocenti} \& {Landolfi}}{{Landi
  degl'Innocenti} \& {Landolfi}}{2004}]{2004ASSL..307.....L}
{Landi degl'Innocenti} E.,  {Landolfi} M.,  2004, {Polarization in Spectral
  Lines}.
Vol.~307 of Astrophysics and Space Science Library, Kluwer Academic Publishers,
  Dordrecht, The Netherlands

\bibitem[\protect\citeauthoryear{{Landstreet} \& {Borra}}{{Landstreet} \&
  {Borra}}{1978}]{1978ApJ...224L...5L}
{Landstreet} J.~D.,  {Borra} E.~F.,  1978, \apjl, 224, L5

\bibitem[\protect\citeauthoryear{{Landstreet} \& {Mathys}}{{Landstreet} \&
  {Mathys}}{2000}]{2000A&A...359..213L}
{Landstreet} J.~D.,  {Mathys} G.,  2000, \aap, 359, 213

\bibitem[\protect\citeauthoryear{{Lanz} \& {Hubeny}}{{Lanz} \&
  {Hubeny}}{2007}]{2007ApJS..169...83L}
{Lanz} T.,  {Hubeny} I.,  2007, \apjs, 169, 83

\bibitem[\protect\citeauthoryear{{Maeder} \& {Meynet}}{{Maeder} \&
  {Meynet}}{2005}]{2005A&A...440.1041M}
{Maeder} A.,  {Meynet} G.,  2005, \aap, 440, 1041

\bibitem[\protect\citeauthoryear{{Manoj}, {Maheswar} \& {Bhatt}}{{Manoj}
  et~al.}{2002}]{2002MNRAS.334..419M}
{Manoj} P.,  {Maheswar} G.,    {Bhatt} H.~C.,  2002, \mnras, 334, 419

\bibitem[\protect\citeauthoryear{{Petit}, {Wade}, {Drissen}, {Montmerle} \&
  {Alecian}}{{Petit} et~al.}{2008}]{2008MNRAS.387L..23P}
{Petit} V.,  {Wade} G.~A.,  {Drissen} L.,  {Montmerle} T.,    {Alecian} E.,
  2008, \mnras, 387, L23

\bibitem[\protect\citeauthoryear{{Preibisch}, {Balega}, {Hofmann}, {Weigelt} \&
  {Zinnecker}}{{Preibisch} et~al.}{1999}]{1999NewA....4..531P}
{Preibisch} T.,  {Balega} Y.,  {Hofmann} K.-H.,  {Weigelt} G.,    {Zinnecker}
  H.,  1999, New Astronomy, 4, 531

\bibitem[\protect\citeauthoryear{{Semel} \& {Li}}{{Semel} \&
  {Li}}{1996}]{1996SolarPhys...164...417S}
{Semel} M.,  {Li} J.,  1996, \solphys, 164, 417

\bibitem[\protect\citeauthoryear{{Sharpless}}{{Sharpless}}{1952}]{1952ApJ...116..251S}
{Sharpless} S.,  1952, \apj, 116, 251

\bibitem[\protect\citeauthoryear{{Shore} \& {Brown}}{{Shore} \&
  {Brown}}{1990}]{1990ApJ...365..665S}
{Shore} S.~N.,  {Brown} D.~N.,  1990, \apj, 365, 665

\bibitem[\protect\citeauthoryear{{Townsend}, {Oksala}, {Cohen}, {Owocki} \&
  {ud-Doula}}{{Townsend} et~al.}{2010}]{2010ApJ...714L.318T}
{Townsend} R.~H.~D.,  {Oksala} M.~E.,  {Cohen} D.~H.,  {Owocki} S.~P.,
  {ud-Doula} A.,  2010, \apjl, 714, L318

\bibitem[\protect\citeauthoryear{{Tuomi}, {Pinfield} \& {Jones}}{{Tuomi}
  et~al.}{2011}]{2011A&A...532A.116T}
{Tuomi} M.,  {Pinfield} D.,    {Jones} H.~R.~A.,  2011, \aap, 532, A116

\bibitem[\protect\citeauthoryear{{ud-Doula}, {Owocki} \& {Townsend}}{{ud-Doula}
  et~al.}{2009}]{2009MNRAS.392.1022U}
{ud-Doula} A.,  {Owocki} S.~P.,    {Townsend} R.~H.~D.,  2009, \mnras, 392,
  1022

\bibitem[\protect\citeauthoryear{{Wade}, {Alecian}, {Bohlender}, {Bouret},
  {Cohen}, {Duez}, {Gagn{\'e}}, {Grunhut}, {Henrichs} \& {Hill}}{{Wade}
  et~al.}{2011}]{2010arXiv1009.3563W}
{Wade} G.~A.,  {Alecian} E.,  {Bohlender} D.~A.,  {Bouret} J.,  {Cohen} D.,
  {Duez} V.,  {Gagn{\'e}} M., et~al.,
  2011, in Active OB stars Vol.~272 of IAU Symposium.
p.~118

\bibitem[\protect\citeauthoryear{{Wade}, {Bagnulo}, {Kochukhov}, {Landstreet},
  {Piskunov} \& {Stift}}{{Wade} et~al.}{2001}]{2001A&A...374..265W}
{Wade} G.~A.,  {Bagnulo} S.,  {Kochukhov} O.,  {Landstreet} J.~D.,  {Piskunov}
  N.,    {Stift} M.~J.,  2001, \aap, 374, 265

\bibitem[\protect\citeauthoryear{{Wade}, {Donati}, {Landstreet} \&
  {Shorlin}}{{Wade} et~al.}{2000a}]{2000MNRAS.313..851W}
{Wade} G.~A.,  {Donati} J.-F.,  {Landstreet} J.~D.,    {Shorlin} S.~L.~S.,
  2000a, \mnras, 313, 851

\bibitem[\protect\citeauthoryear{{Wade}, {Donati}, {Landstreet} \&
  {Shorlin}}{{Wade} et~al.}{2000b}]{2000MNRAS.313..823W}
{Wade} G.~A.,  {Donati} J.-F.,  {Landstreet} J.~D.,    {Shorlin} S.~L.~S.,
  2000b, \mnras, 313, 823

\bibitem[\protect\citeauthoryear{{Wade}, {Howarth}, {Townsend}, {Grunhut},
  {Shultz}, {Bouret}, {Fullerton}, {Marcolino}, {Martins}, {Naz{\'e}},
  {ud-Doula}, {Walborn}, {Donati} \& {the MiMeS Collaboration}}{{Wade}
  et~al.}{2011}]{2011arXiv1106.3008W}
{Wade} G.~A.,  {Howarth} I.~D.,  {Townsend} R.~H.~D.,  {Grunhut} J.~H.,
  {Shultz} M.,  {Bouret} J.-C.,  {Fullerton} A.,  {Marcolino} W.,  {Martins}
  F.,  {Naz{\'e}} Y.,  {ud-Doula} A.,  {Walborn} N.~R.,  {Donati} J.-F.,
  {the MiMeS Collaboration} 2011, \mnras, 416, 3160

\end{thebibliography}
